\documentclass[reprint,aps,pra,superscriptaddress]{revtex4-1}
\usepackage{natbib}
\usepackage{amsmath}
\usepackage{amssymb}
\usepackage{amsthm}
\usepackage{nicefrac}
\usepackage{graphicx}
\usepackage{bm}

\renewcommand{\vec}[1]{\bm{\mathrm{#1}}}

\newcommand{\im}[1]{\operatorname{Im}\left[#1\right]}
\newcommand{\re}[1]{\operatorname{Re}\left[#1\right]}

\newcommand{\tr}[1]{\operatorname{Tr}\left[#1\right]}
\newcommand{\sym}[1]{\operatorname{Sym}\left[#1\right]}
\newcommand{\asym}[1]{\operatorname{Asym}\left[#1\right]}

\newcommand{\unitv}[1]{\hat{\textbf{#1}}}
\newcommand{\uv}[1]{\hat{\textbf{#1}}}
\renewcommand{\vec}[1]{\textbf{#1}}
\newcommand\Ord[1]{\mathcal{O}\left(#1\right)}
\begin{document}

\title{Global $\mathbb{T}$ Operator Bounds on Electromagnetic Scattering: \\
Upper Bounds on Far-Field Cross Sections}
\author{Sean Molesky}
\thanks{Equal contribution}
\author{Pengning Chao}
\thanks{Equal contribution}
\affiliation{Department of Electrical Engineering, Princeton University, Princeton, New Jersey 08544, USA}
\author{Weiliang Jin}
\affiliation{Department of Electrical Engineering, Stanford University, Stanford, California 94305, USA}
\author{Alejandro W. Rodriguez}
\affiliation{Department of Electrical Engineering, Princeton University, Princeton, New Jersey 08544, USA}
\email{arod@princeton.edu}


\begin{abstract}
We present a method based on the scattering $\mathbb{T}$ operator, and
conservation of net real and reactive power, to provide physical
bounds on any electromagnetic design objective that can be framed as a
net radiative emission, scattering or absorption process.  Application
of this approach to planewave scattering from an arbitrarily shaped,
compact body of homogeneous electric susceptibility $\chi$ is found to
predictively quantify and differentiate the relative performance of
dielectric and metallic materials across all optical length scales.
When the size of a device is restricted to be much smaller than the
wavelength (a subwavelength cavity, antenna, nanoparticle, etc.), the
maximum cross section enhancement that may be achieved via material
structuring is found to be much weaker than prior predictions: the
response of strong metals ($\re{\chi} \ll 0$) exhibits a diluted
(homogenized) effective medium scaling $\propto \left|\chi\right| /
\im{\chi}$; below a threshold size inversely proportional to the index
of refraction (consistent with the half-wavelength resonance
condition), the maximum cross section enhancement possible with
dielectrics ($\re{\chi} > 0$) shows the same material dependence as
Rayleigh scattering.  In the limit of a bounding volume much larger
than the wavelength in all dimensions, achievable scattering
interactions asymptote to the geometric area, as predicted by ray
optics.  For representative metal and dielectric materials, geometries
capable of scattering power from an incident plane wave within an
order of magnitude (typically a factor of two) of the bound are
discovered by inverse design.  The basis of the method rests entirely
on scattering theory, and can thus likely be applied to acoustics,
quantum mechanics, and other wave physics.
\end{abstract}
\maketitle

Much of the continuing appeal and challenge of linear 
electromagnetics stems from the same root: given some desired 
objective (enhancing radiation from a quantum emitter~\cite{
eggleston2015optical,aharonovich2016solid,liu2017enhancing,
koenderink2017single,cox2018quantum}, the field intensity in a
photovoltaic~\cite{mokkapati2012nanophotonic,sheng2012light,
ganapati2013light}, the radiative cross section of an antenna~\cite{
shahpari2018fundamental,capek2019optimal,wientjes2014strong}, etc.) 
subject to some practical constraints (material compatibility~\cite{
selvaraja2009subnanometer,briggs2016fully,roberts2018maskless}, 
fabrication tolerances~\cite{lazarov2016length,boutami2019efficient, 
vercruysse2019analytical}, or device size~\cite{ourir2006all,
law2004nanoribbon,caldwell2014sub}), there is currently no method 
for finding uniquely best solutions. 
The associated difficulties are well known~\cite{ahn1998approximate,
polimeridis2014computation,pestourie2018inverse}. 
From plasmonic resonators~\cite{friedler2009solid,santhosh2016vacuum,
chou2018ultracompact} to periodic lattices~\cite{
boroditsky1999spontaneous,lu2016symmetry,yu2017demonstration},
myriad combinations of materials and geometries can be used to
manipulate electromagnetic characteristics (enhancing interactions
with matter~\cite{groblacher2009observation,tang2010optical,
koenderink2015nanophotonics,bliokh2015spin,high2015visible,
flick2018strong}) in extraordinary ways, but net effects are often
similar~\cite{rinnerbauer2013high,dyachenko2016controlling}. 
The wave nature of Maxwell's equations and non-convexity of 
electromagnetic optimizations with respect to material 
susceptibility makes discerning optimal solutions 
challenging~\cite{boyd2004convex,angeris2019computational,
jiang2019simulator}, often leading to the consideration of designs 
that are sensitive to minute structural alterations~\cite{
im2014label,gandhi2019recent}. 
In most situations of practical interest, quantitative relations 
between basic design characteristics (like available volume and 
material response) and achievable performance are not known.

\begin{figure}[t!]
\centering \includegraphics{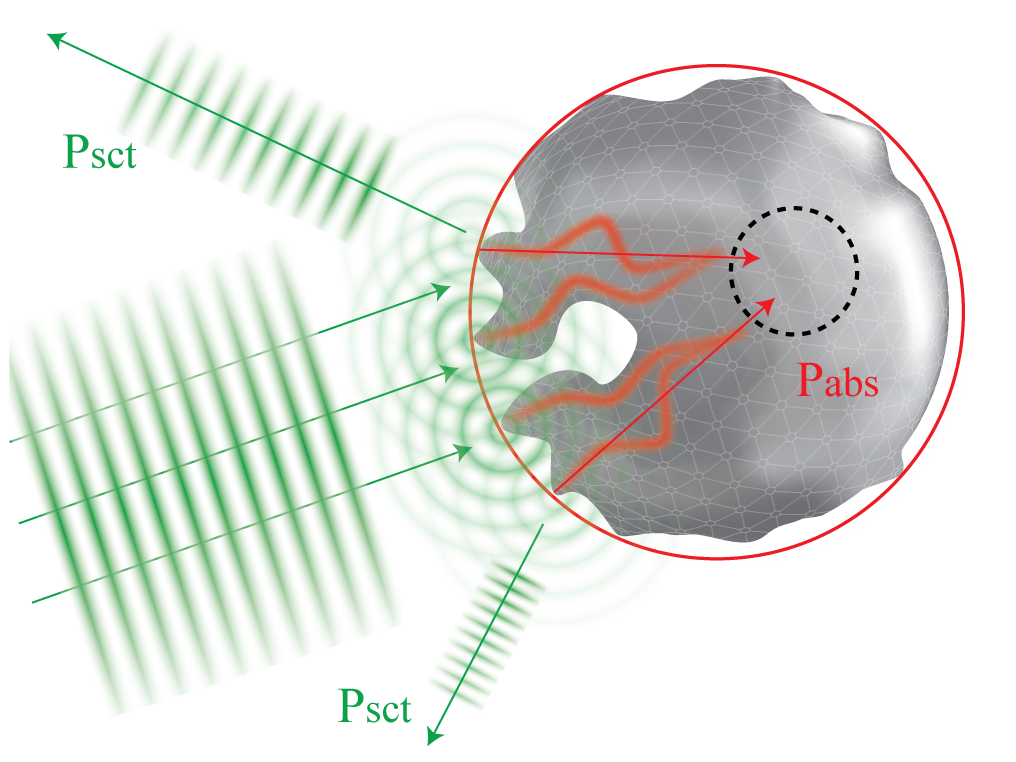}
\vspace{-10 pt}
\caption{\textbf{Schematic of investigation}. 
To what extent does the specification of an electric 
susceptibility ($\chi$) and a confining spatial domain for the 
design of a structured body (optical device), determine the 
maximum absorbed $P_{\text{abs}}$, scattered $P_{\text{sct}}$, and
extinguished (total) $P_{\text{ext}} = P_{\text{sct}} + 
P_{\text{abs}}$ power that can be extracted from a known incident 
field?}
\vspace{0 pt}
\end{figure}

Nevertheless, despite these apparent challenges, computational 
(inverse) design techniques based on local gradients have proven to 
be impressively successful~\cite{jensen2011topology,
molesky2018inverse,lebbe2019robust}, offering substantial 
improvements for applications such as on-chip optical routing~\cite{
frellsen2016topology,su2017inverse,lebbe2019robust}, meta-optics~
\cite{bayati2019inverse,callewaert2018inverse,zhan2019controlling},
nonlinear frequency conversion~\cite{lin2016cavity,sitawarin2018inverse}, and engineered bandgaps~\cite{men2014robust,
meng2018topology}. 
The widespread success of these techniques, and their increasing 
prevalence, begs a number of questions. 
Namely, how far can this progress continue, and, if salient 
limits do exist, can this information be leveraged to either 
facilitate future inverse approaches or define better computational 
objectives. 
Absent benchmarks of what is possible, precise evaluation of the 
merits of any design methodology or algorithm is difficult: failure 
to meet desired application metrics may be caused by issues in the 
selected objectives, the range and type of parameters investigated, 
or the formulation itself. 

Building from similar pragmatic motivations, and basic curiosity, 
prior efforts to elucidate bounds on optical 
interactions, surveyed briefly in Sec.~\ref{priorWork}, have 
provided insights into a diverse collection of topics, including
antennas~\cite{vercruysse2014directional,shahpari2018fundamental,
capek2019optimal}, light trapping~\cite{yablonovitch1982statistical,
siegel1993refractive,yu2012thermodynamic,callahan2012solar,
mokkapati2012nanophotonic,miroshnichenko2018ultimate}, and
optoelectronic~\cite{niv2012near,miller2013photon,xu2015generalized,
liu2016fundamental} devices, and have resulted in improved design 
tools for a range of applications~\cite{munsch2013dielectric,
feichtner2017mode,yang2017low,angeris2019computational}. 
Yet, their domain of meaningful applicability remains fragmented. 
Barring recent computational bounds established by Angeris, Vu{\v{c}}kovi{\'c} and Boyd~\cite{angeris2019computational}, which are limits 
of a different sort, applicability is highly context dependent. 
Relevant arguments largely shift with circumstance~\cite{
molesky2019bounds,molesky2020fundamental}, and even within any 
single setting, attributes widely recognized to affect performance 
(e.g., differences between metallic and dielectric materials, the 
necessity of conserved quantities, required boundaries, minimum 
feature sizes) are frequently unaccounted for, leading to 
unphysical asymptotics and/or bounds many orders of magnitude larger 
than those achieved by state-of-the-art photonic structures, Fig.~1.

In this article, we exploit the requirement of global (net) power
conservation as constructed from the defining relation of the
scattering $\mathbb{T}$ operator, in conjunction with Lagrange
duality, to derive bounds on any electromagnetic objective equivalent
to a net extinction, scattering, or absorption process.  With minor
modifications for cases where one is interested in only a portion of
the output field~\cite{gustafsson2016antenna}, such phenomena
encompass nearly every application mentioned above.  The scheme,
capturing the potential of any and all possible structures under the
fundamental wave limitations contained in Maxwell's equations,
requires only three specified attributes: the material the device will
be made of, the volume it can occupy, and the source (current or
electromagnetic field) that it will interact with.  Directly, the
conservation of real power is seen to set an upper bound on the
magnitude of a system's net polarization response, while the analog of
the optical theorem for reactive power, introducing the polarization
\emph{phase}, is shown to severely restrict the conditions under which
resonant response is attainable, particularly in weak dielectrics
(glasses), leading to significantly tighter limits compared to related
works~\cite{miller2016fundamental,kuang2020maximal}.

The utility of a more robust, methodic, approach for treating
electromagnetic performance limits has been recognized by several
other researchers (especially in the field of radio frequency
antennas~\cite{gustafsson2016antenna,jelinek2016optimal,
  gustafsson2019maximum}), and in particular, concurrent works by
Kuang \textit{et al.}~\cite{kuang2020maximal} and Gustafsson et
al.~\cite{gustafsson2019upper} have converged on the same basic
optimization approach given in Sec.~\ref{formalism}.  Although
Ref.~\cite{kuang2020maximal} considers only the conservation of real
power, and thus overestimates achievable performance for certain
parameter combinations (Figs.~2 and 3), the findings presented in
these articles are excellent complements to our results, further
highlighting the adaptability and utility of Lagrange duality and
scattering theory for predicting possible performance in photonics.
Moreover, during review of this article, another program for
calculating bounds via the self-consistency of the total scattering
field has also been put forward by Trivedi \textit{et al.}~\cite{
  trivedi2020fundamental}.  A brief description of this work, as well
as those cited above, is given in Sec.~\ref{priorWork}.

Application of the technique to compute limits on far-field scattering
cross sections for any object of electric susceptibility $\chi$ that
can be bounded by (contained in) either a ball of radius $R$ or a
periodic film of thickness $t$ interacting with a plane wave of
wavelength $\lambda$, codifies a substantial amount of intuition
pertaining to optical devices.  As $R/\lambda\rightarrow 0$, the
requirement of reactive power conservation means that the energy
transferred between a generated polarization current and its exciting
(incident) field, averaged over the confining volume, can never scale
as the material enhancement factor of $\zeta_{\text{mat}} =
|\chi|^2/\im{\chi}$ introduced and broadly discussed in prior
works~\cite{
  miller2016fundamental,yang2017low,yang2018maximal,michon2019limits,
  molesky2019bounds,venkataram2019fundamental,molesky2020fundamental}.
Instead, for metals ($\re{\chi}\leq -3$), with $\re{\chi} = -3$
corresponding to the localized plasmon-polariton resonance~\cite{
  novotny2012principles} of a spherical nanoparticle, the relative
strength of such interactions cannot surpass, as compared to
$\zeta_{\text{mat}}$, the reduced form factor $3\left|\chi\right|/
\im{\chi}$, consistent with a ``dilution'' of metallic response
implied by homogenized or effective medium perspectives~\cite{
  liu2007description,cai2010optical,jahani2016all}.  When $\re{\chi} >
-3$, even this level of enhancement is not possible.  For dielectrics
($\re{\chi} > 0$), enhancement is limited by the same material
dependence that appears in Rayleigh scattering~\cite{ hulst1981light},
approximately $3 \left|\chi\right| / \left|\chi + 3\right|$.  (In
either case, comparison with cross section limits based on
shape dependent coupled mode or effective medium
models~\cite{hamam2007coupled,merrill1999effective} reveals notable
inexactness.)  After surpassing a wavelength condition inversely
proportional to the index of refraction, the importance of reactive
power is superseded by the conservation of real power, causing
$\re{\chi}$ to have less drastic consequences on the magnitude of
achievable cross sections; for $R \approx \lambda/2$, limits for
dielectrics meet or surpass those of metals for realistic material
loss values, $\im{\chi}$.  In the macroscopic limit of $R\gg\lambda$
or $t \gtrsim \lambda/2$, the selected material plays almost no role
in setting achievable scattering cross sections, other than
determining the level of structuring that would be required, and the
predictions of ray optics (geometric cross sections) emerge.
Additionally, beyond these asymptotic features, for representative
metallic and dielectric material selections, geometries discovered
through topology optimization~\cite{jensen2011topology,
  molesky2018inverse} are shown to come within an order of magnitude
of the bounds for domain sizes varying between $R/\lambda =10^{-3}$
and $R/\lambda=1$, with connate agreement for periodic films,
providing supporting evidence that the bounds are nearly tight and
predictive.

\begin{figure*}[t!]
\centering
\includegraphics[width=2.0\columnwidth]{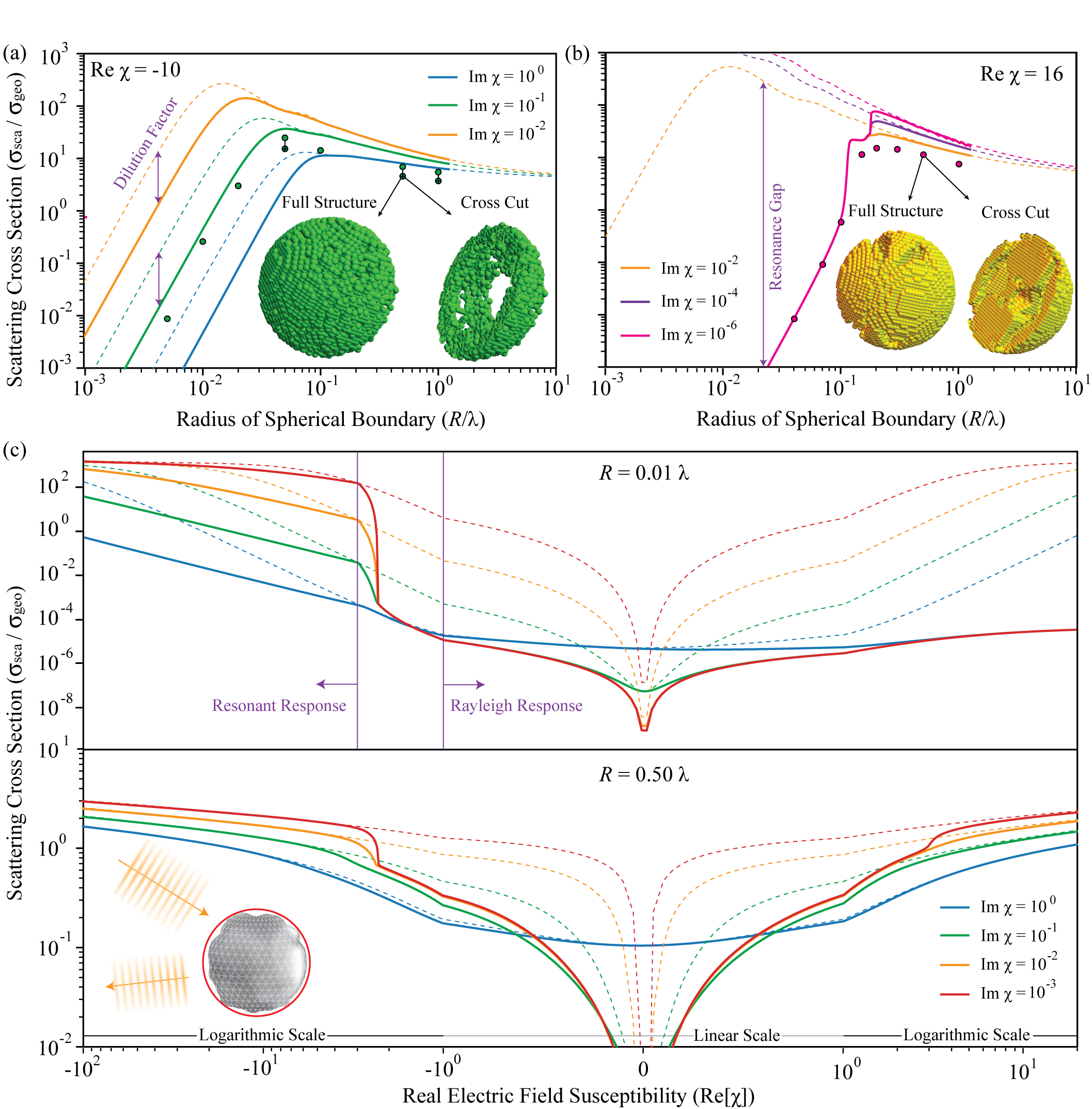}
\vspace{-10 pt}
\caption{\textbf{Scattering cross section bounds for a planewave
incident on a compact object.} 
The four panels show different aspects of scattering cross 
section bounds for any structure of electric susceptibility 
$\chi$  that can be contained within a spherical boundary of 
$R/\lambda$ (e.g., cavities, nanoparticle, etc.). 
Dashed lines result from imposing the conservation of real 
power, as in Ref.~\cite{kuang2020maximal}.
Full lines result from additionally requiring the conservation of
reactive power, as in Ref.~\cite{gustafsson2019upper}. 
As $R\rightarrow 0$, limit values agree with 
\eqref{nonResSmallR} in all cases, and with \eqref{resSmallR} so 
long as $\im{\chi}/\left|\re{\chi}\right|\gtrsim 10^{-4}$. 
Dots in panels (a) and (b) mark scattering cross sections 
achieved in actual geometries discovered by inverse design, for 
$\chi = -10 + i~ 10^{-1}$ and $\chi = 16 + i ~ 10^{-6}$, 
respectively.  
For the metal structure in (a), vertically aligned cross-hatched 
dots result from enforcing binarized permittivity profiles. 
(Binarization does not alter the dielectric results.) 
Two sample structures are shown as insets, with the planewave 
incident from the more solid side of both designs, left side in 
(a), right side in (b), and aligned along the left-right 
symmetry axis. 
Unbounded cross sections are encountered only for fictitious 
metals with vanishing material loss, with the bounds exhibiting 
a weak, sub-logarithmic, divergence as 
$\im{\chi}\rightarrow 0$. 
More practically, cross section enhancements surpassing 
$\approx 200$ should not be expected. 
Descriptions of other major features are given in the 
accompanying text.}
\vspace{-10 pt}
\end{figure*}

\begin{figure*}[t!]
\centering
\includegraphics[width=2.0\columnwidth]{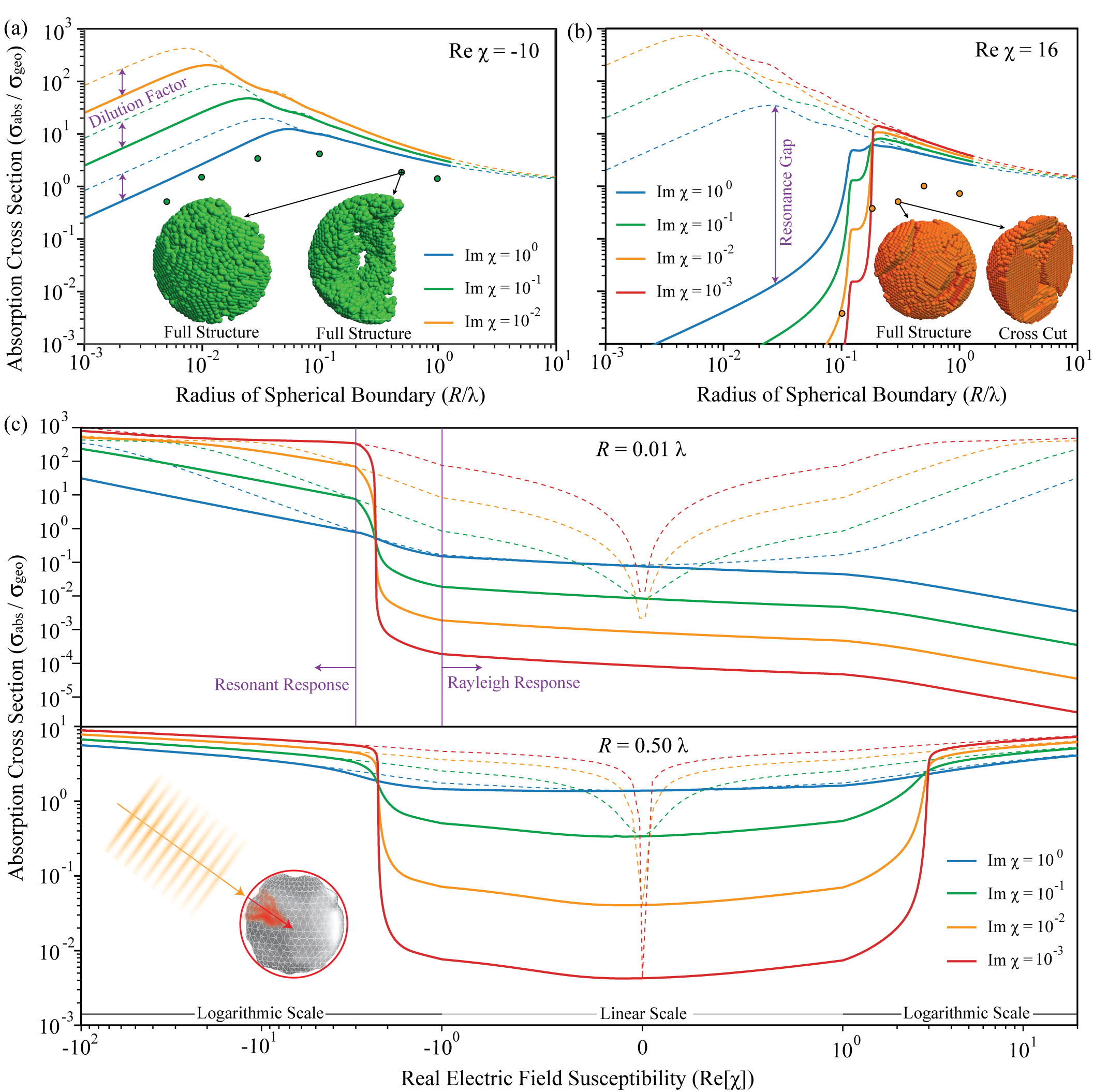}
\vspace{-10 pt}
\caption{\textbf{Absorption cross section bounds for a planewave 
incident on a compact object.} 
The four panels provide analogous information as Fig.~2, but for 
absorbed rather than scattered power. 
Again, the dashed lines represent bounds determined by insisting 
only that real power is conserved, the solid lines result 
from additionally requiring the conservation of reactive power, 
and dots correspond to actual structures discovered by inverse 
design.
For the example metal structure shown in (a), 
the planewave is incident from the solid side of the half 
shell (from the left in the first view and from the back left 
in the second). 
For the example dielectric structure, the planewave is incident 
from the left along the axis of symmetry of the mushroom cap in 
the full view, and from the back left in the cross cut. 
Small radii features are found to be in good agreement with the 
asymptotic predictions of \eqref{nonResSmallR} and 
\eqref{resSmallR} for $\im{\chi}/\left|\re{\chi}\right| 
\gtrsim 10^{-4}$. 
When the confining volume is small, $R/\lambda \lesssim 1/20$, 
the ability of a structured object to absorb radiation is found 
to be substantially weaker than past predictions~\cite{
miller2016fundamental,yang2018maximal,molesky2019bounds}. 
As $R\rightarrow\infty$, the geometric cross section limit 
predicted by ray optics is recovered regardless of the material 
considered. 
Like scattering, these results indicate that absorption cross
section enhancements larger than $\approx 200$ should not be
expected.}
\vspace{-10 pt}
\end{figure*}

These same findings also shed light on a range of fundamental 
questions, such as limitations for light extraction and trapping 
efficiency, and the relative merits of different materials for 
particular applications~\cite{liu2016fundamental,jahani2016all,
staude2019all}, and provide a much more quantitative perspective on 
which aspects of a design are most critical to device performance 
than prior analyses. 
Shortly, we foresee extensions of the framework to embedded sources 
and user defined design (containing) volumes as providing a means of 
formalizing, comparing, and contrasting existing paradigms within 
photonics, revealing limitations and trade offs in a number of 
technologically prescient areas (e.g., the radiative efficiency of 
quantum emitters~\cite{lu2017dynamically,davoyan2018quantum,
crook2020purcell}, high quality factor cavities~\cite{
lin2016enhanced,liu2017quantum,wang2018maximizing}, optical forces~
\cite{venkataram2020casimir}, luminescence~\cite{zalogina2018purcell,
valenta2019nearly} and fluorescence~\cite{li2017plasmon,
simovski2019point}).

The article is divided into four main sections.
Section~\ref{formalism} begins with an overview of the $\mathbb{T}$
operator relations governing absorption, scattering and radiative
processes, followed by a statement of the wave constraints and
relaxations that are explored in the rest of the manuscript. From
these preliminaries, the calculations of limits is then cast in the
language of optimization theory, and a solution in terms of the
Lagrangian dual is given. In Sec.~\ref{priorWork}, a brief
examination of the similarities and differences of this approach with
current art is supplied. Next, in Sec.~\ref{toyBounds}, basic
computational mechanics are examined, and the quasistatic
$R/\lambda\rightarrow 0$ scaling factors quoted elsewhere are derived
based on analytically tractable single-channel asymptotics.
Finally, Sec.~\ref{applications} provides sample applications of the 
method as described above. 
This is likely the section of the text of most relevance to the 
majority of readers. 
Although only single frequency examples are
given, broad bandwidth objectives should present no major
hurdles~\cite{liang2013formulation,shim2019fundamental}.  Further
technical details, necessary only for reproducing our results, appear
in Sec.~\ref{appendix}.  Because the approach relies exclusively on
the validity of scattering theory, it is likely that counterparts of
the presented findings exist in acoustics, quantum mechanics, and any
other wave physics.

\section{Formalism}\label{formalism}

The key to the approach of this article rests on the use of partial 
relaxations~\cite{pardalos2013handbook}. 
Past electromagnetic limits, including our own work, have been 
predominately formulated by placing bounds on the individual 
physical quantities entering an objective and then deducing a total 
bound by composing the component limits~\cite{miller2007fundamental,
miller2015shape}. 
We begin, alternatively, with the fundamental scattering relation 
that any physical system must satisfy, derive algebraic consequences 
of this relations (e.g. energy conservation) and then suppose a 
subset of theses conclusions as constraints on an otherwise 
abstract optimization. 
This general procedure is detailed below. 
The formulas given in the \emph{Power Objectives} and 
\emph{Scattering Constraints} subsections are exact. 
Relaxations (omissions of certain physical requirements) begin 
only after the \emph{Relaxation and Optimization} heading. 

\subsection{Power Objectives}

Considerations of power transfer in electromagnetics typically belong
to one of two categories: initial flux problems, wherein power is
drawn from an incident electromagnetic field, and initial source
problems, wherein power is drawn from a predefined current 
excitation.
Initial flux problems are typical in scattering theory, and as such,
our nomenclature follows essentially from this area~\cite{
tsang2004scattering}. 
That is, we will denote the \emph{initial (incident, given, or bare) 
field} with an $i$ superscript (either $\left|\textbf{E}^{i}\right>$ 
or $\left|\textbf{J}^{i}\right>$) and the \emph{total (or dressed) 
field} with a $t$ superscript. 
For a pair of initial and total quantities referring to the same 
underlying field, the \emph{scattered field}, $s$ superscript, is 
defined as the difference $\left|\textbf{F}^{s}\right> = \left|
\textbf{F}^{t}\right> - \left|\textbf{F}^{i}\right>.$ 
There is a certain appeal to transforming one of these two classes 
of problem into the other via equivalent fields. 
However, due to the additional back action that can occur in initial 
source problems, in our experience a unified framework promotes 
logical slips. 
For this reason, the total polarization field of an initial flux 
problem (or total electromagnetic field of an initial source 
problem) will be referred to as a \emph{generated field}, $g$ 
superscript. 
With this notation, scattering theory for initial flux and source 
problems consists of the following relations:
\begin{align}
\left|\textbf{J}^{g}\right> &= -\frac{i k}{Z}\mathbb{V}\left|
\textbf{E}^{t}\right>, 
& \left|\textbf{E}^{t}\right> 
&=\mathbb{V}^{-1}\mathbb{T}\left|\textbf{E}^{i}\right>,\nonumber
\\ 
\left|\textbf{E}^{t}\right> &= \left|\textbf{E}^{i}\right> 
+\frac{iZ}{k}\mathbb{G}^{\text{0}}\left|\textbf{J}^{g}\right>,
&\left|\textbf{E}^{s}\right> &=
\frac{iZ}{k}\mathbb{G}^{\text{0}}\left|\textbf{J}^{g}\right>,
\label{initFlux}
\end{align}
\begin{align}
\left|\textbf{E}^{g}\right> &=
\frac{iZ}{k}\mathbb{G}^{\text{0}}\left|\textbf{J}^{t}\right>, &
\left|\textbf{J}^{t}\right> &=
\mathbb{T}\mathbb{V}^{-1}\left|\textbf{J}^{i}\right>,
\nonumber\\ 
\left|\textbf{J}^{t}\right>&= \left|\textbf{J}^{i}\right> -
\frac{ik}{Z}\mathbb{V}\left|\textbf{E}^{g}\right>, 
&\left|\textbf{J}^{s}\right> &= -\frac{ik}{Z} \mathbb{V} \left|
\textbf{E}^{g}\right>.
\label{initSource}
\end{align}
Here and throughout, $\mathbb{G}^{\text{0}}$ marks the \emph{
background} or \emph{environmental} Green function, which may or 
may not be vacuum~\cite{novotny2012principles}. 
The $\mathbb{V}$ operator refers to the scattering
potential (susceptibility) relative to this background (whatever
material was not included when $\mathbb{G}^{\text{0}}$ was computed),
and $\left|\textbf{E}^{i}\right>$ and $\left|\textbf{J}^{i}\right>$
are similarly defined as initial fields (solutions) in the 
background. 
The remaining quantities in \eqref{initFlux} and \eqref{initSource} 
are the impedance of free space $Z$, the wavenumber $k = 2\pi/
\lambda$ (with $\lambda$ the wavelength), and the $\mathbb{T}$ 
operator, defined by the formal equality $\mathbb{I}=(\mathbb{V}^{-1}
-\mathbb{G}^{\text{0}})\mathbb{T}$~\cite{molesky2020fundamental}.

The three primary operator forms for energy transfer in an initial
flux problem are the \emph{extracted power},
\begin{equation}
P_{\text{flx}}^{\text{ext}} = 
\frac{1}{2}\re{\left<\textbf{E}^{i}|\textbf{J}^{g}\right>} = 
\frac{k}{2Z}\tr{\mathbb{S}_{E}\asym{\mathbb{T}}},
\label{fExt}
\end{equation}
the \emph{absorbed power},
\begin{align}
P_{\text{flx}}^{\text{abs}} &= 
\frac{1}{2}\re{\left<\textbf{E}^{t}|\textbf{J}^{g}\right>} = 
\frac{k}{2Z}\tr{\mathbb{S}_{E}\left(\mathbb{T}^{\dagger}
\asym{\mathbb{V}^{-1\dagger}}\mathbb{T}\right)},
\nonumber \\
&= \frac{k}{2Z}\tr{\mathbb{S}_{E}
\left(\asym{\mathbb{T}} -\mathbb{T}^{\dagger}
\asym{\mathbb{G}^{\text{0}}}\mathbb{T}\right)},
\label{fAbs}
\end{align}
and the \emph{scattered power},
\begin{align}
P_{\text{flx}}^{\text{sct}} &= 
-\frac{1}{2}\re{\left<\textbf{E}^{s}|\textbf{J}^{g}\right>} = 
\frac{k}{2Z}\tr{\mathbb{S}_{E} \mathbb{T}^{\dagger}
\asym{\mathbb{G}^{\text{0}}}\mathbb{T}} \nonumber \\
&=\frac{k}{2Z}\tr{\mathbb{S}_{E} 
\left(\asym{\mathbb{T}} - \mathbb{T}^{\dagger}
\asym{\mathbb{V}^{-1\dagger}}\mathbb{T}\right)};
\label{fSct}
\end{align}
with $\mathbb{S}_{E} =\left|\textbf{E}^{i}\right>\left<\textbf{E}^{i}
\right|$ denoting projection of the corresponding operators onto the 
incident fields.
Reciprocally, the three principal forms 
characterizing power flow from an initial current excitation are the 
\emph{extracted power},
\begin{align}
P_{\text{src}}^{\text{ext}} &= -\frac{1}{2}
\re{\left< \textbf{J}^{i} | \textbf{E}^{g} \right>} 
\nonumber \\
&= \frac{Z}{2k}\tr{\mathbb{S}_{J}
\left(\asym{\mathbb{V}^{-1\dagger}} + 
\asym{\mathbb{V}^{-1}\mathbb{T}\mathbb{V}^{-1}}\right)},
\nonumber \\
&=\frac{Z}{2k}
\tr{\mathbb{S}_{J}\left(
\asym{\mathbb{G}^{\text{0}}} + 
\asym{\mathbb{G}^{\text{0}}\mathbb{T}\mathbb{G}^{\text{0}}}\right)}
\label{cExt}
\end{align}
the \emph{radiated power},
\begin{align}
P_{\text{src}}^{\text{rad}} &= -\frac{1}{2}
\re{\left<\textbf{J}^{t}|\textbf{E}^{g}\right>} 
\nonumber \\ 
&= \frac{Z}{2k}\tr{\mathbb{S}_{J}\left(
\mathbb{V}^{-1\dagger}\mathbb{T}^{\dagger}
\asym{\mathbb{G}^{\text{0}}}\mathbb{T}\mathbb{V}^{-1}\right)} 
\nonumber \\
&= \frac{Z}{2k}\operatorname{Tr}\big[\mathbb{S}_{J}\mathbb{V}^{-1\dagger}
\left(\asym{\mathbb{T}}\right. -
\nonumber \\ 
&\hspace{0.55in}\left.\mathbb{T}^{\dagger}
\asym{\mathbb{V}^{-1\dagger}}\mathbb{T}\right)\mathbb{V}^{-1}\big]
\nonumber \\
&= \frac{Z}{2k}\operatorname{Tr}\big[\mathbb{S}_{J}
\big(\asym{\mathbb{G}^{\text{0}}} + 
\asym{\mathbb{G}^{\text{0}}\mathbb{T}\mathbb{G}^{\text{0}}}- 
\nonumber\\
&\hspace{0.60in}\mathbb{G}^{\text{0}\dagger}
\mathbb{T}^{\dagger}\asym{\mathbb{V}^{-1\dagger}}
\mathbb{T}\mathbb{G}^{\text{0}}\big)\big],
\label{cRad}
\end{align}
and the \emph{material (loss) power},
\begin{align}
P_{\text{src}}^{\text{mat}} &= 
\frac{1}{2}\re{\left<\textbf{J}^{s}|\textbf{E}^{g}\right>} 
\nonumber \\ &= \frac{Z}{2k}\tr{
\mathbb{S}_{J}\left(\mathbb{G}^{\text{0}\dagger}
\mathbb{T}^{\dagger}
\asym{\mathbb{V}^{-1\dagger}}\mathbb{T}\mathbb{G}^{\text{0}}\right)} 
\nonumber \\
&= \frac{Z}{2k}\operatorname{Tr}\big[\mathbb{S}_{J} 
\left(\mathbb{G}^{\text{0}\dagger}\asym{\mathbb{T}}
\mathbb{G}^{\text{0}} -\right.
\nonumber \\ 
&\hspace{0.55in}\left.\mathbb{G}^{\text{0}\dagger}
\mathbb{T}^{\dagger}\asym{\mathbb{G}^{\text{0}}}
\mathbb{T}\mathbb{G}^{\text{0}}\right)\big],
\label{cMat}
\end{align}
with $\mathbb{S}_{J} = \left|\textbf{J}^{i}\right>\left<
\textbf{J}^{i}\right|$ denoting projection of the corresponding 
operators onto the initial current sources. 
The naming of the final two forms, which appear less frequently than 
the other four, follows from the observation that once the total 
source $\left|\textbf{J}^{t}\right>$ is determined the corresponding 
electromagnetic field is generated exclusively via the background 
Green's function. 
Hence, the energy transfer dynamics of a total source are exactly 
those of a special ``free'' current distribution. 
Because the only pathway for power to flow from a current source in 
free space (or lossless background) is radiative emission, 
$P^{\text{rad}}_{\text{src}}$ must be interpreted in this 
way---energy transfer into the source free solutions of the background---and similarly, $P^{\text{mat}}_{\text{src}}$ must be
equated with loss into the scatterer. 
This reversal of forms and physics (compared absorption and 
scattering) is sensible from the perspective of field conversion. 
Absorption is the conversion of an electromagnetic field into a 
current, and radiative emission the conversion of a current into an 
electromagnetic field. 
Scattering in an initial flux setting is the creation of a new field
of the same type, as is material loss in an initial source setting.

Note, however, that there is a caveat to these interpretations. 
$\mathbb{G}^{\text{0}}$, as we have defined it, accounts for the 
entire electromagnetic background. 
As such, in non vacuum cases, any polarization current 
associated with a background solution will not appear in any $\left|
\textbf{J}\right>$ field, generally comingling the corresponding 
physical interpretation of the various power quantities.  
For instance, as $\asym{\mathbb{G}^{\text{0}}}$ describes power flow 
into the full homogeneous solutions of Maxwell's equations, if the 
environment for which $\mathbb{G}^{\text{0}}$ is determined contains 
absorptive material then $\asym{\mathbb{G}^{\text{0}}}$ does not 
represent radiation. 
Implied meaning can be restored by appropriate alterations; but, as 
this point will be treated in an upcoming work, for the moment we 
will simply accept it as a limitation for our study. 

Setting such possibilities aside, the equivalence of \eqref{cRad} 
with radiative emission is also supported both by 
the analogy between its operator form and that of the scattered 
power, and by a direct calculation for thermal (randomly 
fluctuating) currents~\cite{polimeridis2015fluctuating}. 
By the fluctuation--dissipation theorem $\langle \left|\textbf{J}^{i}
\right>\left< \textbf{J}^{i}\right|\rangle_{\text{th}} = 4 k\Pi\left(
\omega,T\right)\asym{\mathbb{V}} / \left(\pi Z\right)$, and so
\begin{align}
P_{\text{th}} &= \frac{Z}{2 k}\langle
\im{\left<\textbf{J}^{i}\right| \mathbb{V}^{-1\dagger} 
\mathbb{T}^{\dagger} \mathbb{G}^{\text{0}}\mathbb{T}\mathbb{V}^{-1}
\left|\textbf{J}^{i}\right>}\rangle_{\text{ther}} 
\nonumber \\
&= \frac{Z}{2k}\im{\tr{\langle\left|\textbf{J}^{i}\right> 
\left<\textbf{J}^{i}\right|\rangle_{\text{ther}}
\mathbb{V}^{-1\dagger}\mathbb{T}^{\dagger} 
\mathbb{G}^{\text{0}} \mathbb{T} \mathbb{V}^{-1}}}
\nonumber \\
&= \frac{2~\Pi\left(\omega,T\right)}{\pi}
\tr{\asym{\asym{\mathbb{V}^{-1\dagger}}
\mathbb{T}^{\dagger}\mathbb{G}^{\text{0}}\mathbb{T}}}
\nonumber \\
&= \frac{2~\Pi\left(\omega,T\right)}{\pi}
\operatorname{Tr}\big[\left(\asym{\mathbb{T}} -
\mathbb{T}\asym{\mathbb{G}^{\text{0}}}\mathbb{T}^{\dagger}\right)\times
\nonumber
\\ &\hspace{1.0in}\asym{\mathbb{G}^{\text{0}}}\big].
\label{themralEmission}
\end{align}
The final line above is precisely what we have derived in Ref.~\cite{
molesky2019bounds} from the perspective of integrate absorption.

\subsection{Scattering Constraints}

As supported by the previous subsection, any 
quantity in electromagnetics can be described by combinations 
of $\mathbb{G}^{\text{0}}$, $\mathbb{V}$, $\mathbb{T}$, and 
projection operators. 
The basis of this reality rest on the fact that a defining 
relation for $\mathbb{T}$, supposing $\mathbb{G}^{\text{0}}$ and $
\mathbb{V}$ are known, is abstractly equivalent to complete 
knowledge of the electromagnetic system~\cite{tsang2004scattering}. 
Thus, like Maxwell's equations, any facet of classical 
electromagnetics, beyond the definitions of $\mathbb{G}^{\text{0}}$ 
and $\mathbb{V}$, must be derivable from the definition of the 
$\mathbb{T}$ operator~\cite{tsang2004scattering,molesky2019bounds}
\begin{equation}
\mathbb{I}_{ss} = \left(\mathbb{V}^{-1}_{ss} - 
\mathbb{G}^{\text{0}}_{ss}\right)\mathbb{T}_{ss}.
\label{Tformal}
\end{equation}
(The $s$ subscript explicitly marks the domain and codomain
of each operator as begin restricted to the scattering object, 
as opposed to the background.) 
To derive constraints, we will focus on the equivalent relation 
\begin{align}
\mathbb{T}_{ss} = \mathbb{T}_{ss}^{\dagger}
\left(\mathbb{V}^{\dagger -1}_{ss}-\mathbb{G}^{\text{0}\dagger}_{ss}\right)
\mathbb{T}_{ss}.
\label{optFunda1}
\end{align}
As both $\mathbb{T}_{ss}$ and its Hermitian conjugate $
\mathbb{T}_{ss}^{\dagger}$ have support only within the volume of 
the scatterer, in this form, the domain and codomains of 
$\mathbb{V}^{\dagger-1}$ and $\mathbb{G}^{\text{0}\dagger}$ are 
unimportant. 
For any projection operator $\left(\forall n\in
\mathbb{N}\right)~\mathbb{P} = \mathbb{P}^{n}$, and so the geometric 
description of the scatterer contained in $\mathbb{V}^{
\dagger -1}_{ss}$ and $\mathbb{G}^{\text{0}\dagger}_{ss}$ are 
unnecessary since $\mathbb{T}_{ss}^{\dagger}\mathbb{V}^{
\dagger -1}_{ss}\mathbb{T}_{ss} = \mathbb{T}_{ss}^{\dagger}
\mathbb{V}^{\dagger -1}\mathbb{T}_{ss}$ and $\mathbb{T}_{ss}^{
\dagger}\mathbb{G}^{0\dagger}_{ss} \mathbb{T}_{ss} = 
\mathbb{T}_{ss}^{\dagger}\mathbb{G}^{\text{0}\dagger}
\mathbb{T}_{ss}$. 
This makes 
\begin{align}
\mathbb{T}_{ss} = \mathbb{T}_{ss}^{\dagger}
\left(\mathbb{V}^{\dagger -1}-\mathbb{G}^{\text{0}\dagger}\right)
\mathbb{T}_{ss},
\label{optFunda2}
\end{align}
equivalent to \eqref{optFunda1}, with the scattering potential 
$\mathbb{V}^{\dagger-1}$ and Green's function 
$\mathbb{G}^{\text{0}\dagger}$ filling whatever volume we would 
like to consider. 
Taking 
\begin{equation}
\mathbb{U} = \mathbb{V}^{\dagger-1}-\mathbb{G}^{\text{0}
\dagger}, 
\label{uDef}
\end{equation} 
so that $\asym{\mathbb{U}}$ is positive definite, treating the 
symmetric (Hermitian) and anti-symmetric (skew Hermitian) parts of 
\eqref{optFunda2} separately then gives 
\begin{align}
&\sym{\mathbb{T}_{ss}} = \mathbb{T}_{ss}^{\dagger}\sym{\mathbb{U}}\mathbb{T}_{ss},
\label{symOpt} 
\\
&\asym{\mathbb{T}_{ss}} = \mathbb{T}_{ss}^{\dagger}\asym{\mathbb{U}}\mathbb{T}_{ss}.
\label{asymOpt}
\end{align}
The constraints used to generate the cross section bounds shown in
Sec.~\ref{applications} follow directly from \eqref{symOpt} and
\eqref{asymOpt} under the relaxation described in the next section.

As recently noted in Refs.~\cite{molesky2019bounds,
molesky2020fundamental,venkataram2019fundamental,gustafsson2019upper}
, \eqref{symOpt} and \eqref{asymOpt} contain a surprising amount of 
physics. 
Taken together, these relations give a full algebraic 
characterization of power conservation~\cite{jackson1999classical,
gustafsson2019upper}, with \eqref{symOpt} representing the 
conservation of reactive power and \eqref{asymOpt} the conservation 
of real power. 
(The $\mathbb{T}^{\dagger}_{ss}\sym{\mathbb{U}}\mathbb{T}_{ss}$ 
piece of \eqref{symOpt} produces the difference of the 
magnetic and electric energies for any incident electric field~
\cite{harrington1971theory}.) 
Because both real and imaginary response are captured, when these 
two constraints are employed there are requirements that must be 
satisfied on both the magnitude and phase of any potential resonance.

\subsection{Relaxations and Optimization}

For the single source problems of concern to this article, it is
simplest to work from the perspective of the image field resulting 
from the action of $\mathbb{T}_{ss}$ on a given source $\left|
\textbf{S}^{\left(1\right)}\right>$, $\mathbb{T}_{ss}\left|
\textbf{S}^{\left(1\right)}\right> \mapsto \left|\textbf{T}\right>$. 
A bound in this setting amounts to a global maximization of one of 
the six power transfer objectives, \eqref{fExt}--\eqref{cMat}, 
taking $\left|\textbf{T}\right>$ and a known linear functional 
$\left<\textbf{S}^{\left(2\right)}\right|$ as arguments, subject to 
satisfaction of \eqref{symOpt} and \eqref{asymOpt} as applied to the 
source and its image. 
So long as the known fields are not altered at previously included 
locations by expanding the domain, this procedure leads to domain 
monotonic growth: if $\left|\textbf{S}^{\left(1\right)}\right>$ and 
$\left|\textbf{T}\right>$ satisfy all constraints on some 
sub-domain, then these same vectors will also satisfy the 
constraints if they are embedded (included without alteration) into 
a larger domain. 
Because the value of any power objective is similarly unaffected by 
inclusion, the global maximum of a larger domain will thus always be 
larger than the global maximum of a smaller domain. 

The above view also underlies the central relaxation, persisting 
throughout the remainder of the article, that makes global optimization over all structuring alternatives possible. 
For any true $\mathbb{T}$ operator, nonzero polarization currents 
can exist only at spatial points lying within the scattering object. 
This fact will never be truly enforced on the image of the source 
resulting from the action of $\mathbb{T}$ ($\left|\textbf{T}
\right>$), alleviating the need for an explicit geometric description of the scatterer. 
Rather, $\left|\textbf{T}\right>$ will be considered simply as an 
unknown vector field confined to some predefined design domain.  
(For instance, in the examples of Sec.~\ref{toyBounds} and Sec.
~\ref{applications} the encompassing design domain is taken to be 
ball of radius $R$.)  
Therefore, when a bound is found, it must necessarily apply to any 
possible structure that can be contained in the given region, 
as the freedom of choosing different device geometries is already 
explored by optimizing over the $\left|\textbf{T}\right>$ field. 
As illustration, through this relaxation of structural information 
and the monotonicity property, a bound for a cuboid is both a bound 
for any device that could fit inside the cuboid, no matter how 
exotic, and for any sub-design region that could be included inside 
the cuboid, be it a needle, bounded fractal, or a disconnected 
collection of Mie scatterers. 

With this easing of true physical requirements in mind, scattering 
operator bounds for any of \eqref{fExt}--\eqref{cMat} are equated 
with an optimization problem on $\left|\textbf{T}\right>$: 
\begin{align}
&\text{max}~\mathcal{O} = \sum_{\ell}
\im{\left<\textbf{S}^{\left(2\right)}_{\ell}
\Big|\textbf{T}_{\ell}\right>} - 
\left<\textbf{T}_{\ell}\Big|\mathbb{O}_{\ell}
\Big|\textbf{T}_{\ell}\right>
\nonumber \\
&\text{such that} \\
\nonumber
&\mathcal{C}_{\zeta} = \sum_{\ell}\im{\left<
\textbf{S}^{\left(1\right)}_{\ell}\Big|
\textbf{T}_{\ell}\right>} - \left<\textbf{T}_{\ell}
\Big|\asym{\mathbb{U}_{\ell}}
\Big|\textbf{T}_{\ell}\right> = 0,
\nonumber \\
&\mathcal{C}_{\gamma} = \sum_{\ell}
\re{\left<\textbf{S}^{\left(1\right)}_{\ell}
\Big|\textbf{T}_{\ell}\right>} - 
\left<\textbf{T}_{\ell}\Big|\sym{\mathbb{U}_{\ell}}
\Big|\textbf{T}_{\ell}\right> = 0.
\nonumber \\
\label{optProb}
\end{align}
The corresponding Lagrangian is given by
\begin{align}
\mathcal{L} = &\sum_{\ell}\im{\left<
\textbf{S}^{\left(2\right)}_{\ell}\Big|
\textbf{T}_{\ell}\right>} - 
\left<\textbf{T}_{\ell}\Big|\mathbb{O}_{\ell}\Big|
\textbf{T}_{\ell}\right> +
\nonumber \\
&\zeta\left(\im{\left<\textbf{S}^{\left(1\right)}_{\ell}
\Big|\textbf{T}_{\ell}\right>} - \left<\textbf{T}_{\ell}
\Big|\asym{\mathbb{U}_{\ell}}\Big|\textbf{T}_{\ell}\right>
\right) + 
\nonumber \\
&\gamma\left(\re{\left<\textbf{S}^{\left(1\right)}_{\ell}
\Big|\textbf{T}_{\ell}\right>} - \left<\textbf{T}_{\ell}
\Big|\sym{\mathbb{U}_{\ell}}\Big|\textbf{T}_{\ell}\right>
\right).
\label{Lgen}
\end{align}
To help decompose later analysis, in \eqref{optProb} and 
\eqref{Lgen} the subscript $\ell$ has been introduced to stand for a 
\emph{family} of $j$ indices coupled together by $\mathbb{U}$ 
($\mathbb{U}_{\ell} = \mathbb{V}^{\dagger-1}_{\ell}-\mathbb{G}^{
\text{0}\dagger}_{\ell}$) in some complete basis $\left\{\left|
\textbf{G}_{\ell,j}\right>\right\}$ for vector fields in domain, 
as would likely occur in any numerical solution approach. 
For the spherically bounded examples examined in 
Sec.~\ref{toyBounds} and Sec.~\ref{applications} $\ell$ is the 
familiar angular the momentum number, for a periodic films $\ell$ is 
an in-plane wave vector. 
(Generally, $\ell$ can be thought of as the $\ell$th block, 
invariant subspace, in the representation $\left<\textbf{G}_{\ell,i}
\right|\mathbb{U}\left|\textbf{G}_{\ell,j}\right>$, and $j$ is used 
as a subindex.)  
Due to the further equivalence of the angular index and the 
radiation modes of a spherical domain, as well as relations with 
existing literature ~\cite{molesky2019bounds,miller2000communicating,
miller2007fundamental,pendry1983quantum}, we will also occasionally 
refer to $\ell$ families as \emph{channels} (as a shorthand for 
radiation channels).
The constraints $\mathcal{C}_{\zeta}$ and $\mathcal{C}_{\gamma}$
are determined by applying \eqref{symOpt} and \eqref{asymOpt} to
$\left\{\left<\textbf{S}^{\left(1\right)}\right|,~\left|
\textbf{S}^{\left(1\right)}\right>\right\}$, and forgetting any
information related to the geometry of the scatterer. 
$\mathbb{O}$ is either $\asym{\mathbb{G}^{\text{0}}}$, 
$\asym{\mathbb{V}^{\dagger-1}}$, or $\textbf{0}$, depending on 
whether the problem is absorption/material loss, 
scattering/radiation or extracted power from a field. 
As exemplified in Sec.~\ref{toyBounds} and illustrated in 
Sec.~\ref{applications}, the necessity of conserving reactive power
imparted by the symmetric $\gamma$ constraint is crucial for 
accurately anticipating how a particular choice of material and 
domain influences whether or not a family can achieve resonant 
response. 

For all cases except extracted and radiated power from an external
unpolarizable source, $\left<\textbf{S}^{\left(2\right)}\right| =
\left<\textbf{S}^{\left(1\right)}\right|$. 
In these instances, although \eqref{cExt} and \eqref{cRad} show that 
the power quantities of interest can be cast in a form similar 
to the corresponding initial flux problems, the inclusion of the 
second source image is necessary. 
If an unpolarizable source is taken to lie outside the domain being 
optimized, $\mathbb{V}^{-1}$ is defined only as a limit (tending to 
infinity). 
Once this limit is taken, the $\mathbb{G}^{\text{0}}$ based 
expressions (final forms) for the extracted and radiated power 
result, which include the introduction of the field $\left|
\textbf{S}^{\left(2\right)}\right> = \mathbb{G}^{0*}_{de}\left|
\textbf{J}^{i}\right>$ to the objective due to the 
$\asym{\mathbb{G}^{\text{0}}\mathbb{T}\mathbb{G}^{\text{0}}}$ term. 
(With $e$ denoting the external space of the emitter and $d$ the 
optimization domain.) 
These differences amount to the introduction of cross terms 
describing the interference of the fields generated by the bare and 
induced currents that are no longer inherently accounted for by 
the scattered currents at the location of the source (multiple 
scattering and back action). 
Still, the form of these problems remains like \eqref{Lgen} up to 
the addition of an unalterable background contribution of 
$\tr{\mathbb{S}_{J}\asym{\mathbb{G}^{\text{0}}}}$. 

\subsection{Solution via Duality}

To solve \eqref{optProb}, we make use of the following lemma, 
commonly referred to as Lagrange duality~\cite{boyd2004convex}. 
(Lagrange duality is closely associated with the alternating 
direction method of multipliers~\cite{lu2010inverse,
boyd2011distributed,lu2013nanophotonic} often used for solving 
multiply constrained optimization problems.)
\\ \\ 
\emph{Lagrange Duality.} 
Take $\mathcal{O}$, $\left(\forall j\in J\right)~\mathcal{E}_{j}$ 
and $\left(\forall k\in K\right)~\mathcal{I}_{k}$ to be 
differentiable real valued functions on subsets of $\mathbb{R}^{n}$ 
defining a well-posed optimization problem
\begin{align}
&\text{max}~\mathcal{O}\left(\textbf{x}\right) ~~ 
\left(\textbf{x}\in\mathbb{R}^{n}\right)
\nonumber\\
&\text{such that }\left(\forall k\in K\right)~
\mathcal{I}_{k}\left(\textbf{x}\right) \geq 0~
\&~\left(\forall j\in J\right)~
\mathcal{E}_{j}\left(\textbf{x}\right) = 0.
\nonumber
\end{align} 
Let $m_{*}$ be the corresponding maximum value and let $\mathcal{D}$ 
be the domain on which all functions are simultaneously defined, 
which includes the subset $\mathcal{D}_{*}$ on which all constraints 
are satisfied. 
Take $\mathcal{L} = \mathcal{O}\left(\textbf{x}\right) + \sum_{j
\in J}\lambda_{j}\mathcal{E}_{j}\left(\textbf{x}\right) + \sum_{k
\in K}\nu_{k}\mathcal{I}_{k}\left(\textbf{x}\right)$ to be the
Lagrangian of the optimization. 
For any collection of values of $\left\{\lambda_{j}\right\}_{J}$ and 
$\left\{\nu_{k}\geq 0 \right\}_{K}$
$$m_{*}\leq \text{max}_{\mathcal{D}}~\mathcal{L}\left(\textbf{x},
\left\{\lambda_{j}\right\}_{J},\left\{\nu_{k}\right\}_{K}\right).$$ 
Additionally, the function $\text{max}_{\mathcal{D}}~\mathcal{L}$ is 
convex, and, if a set $\left\{\left\{\lambda_{j}\right\}_{J}, \left
\{\nu_{k}\right\}_{K}\right\}$ (with $\left(\forall k\in K\right) 
\nu_{k}\geq 0$) minimizing $\text{max}_{\mathcal{D}}~\mathcal{L}$ is 
found such that $\left(\forall j\in J\right)~\mathcal{E}_{j}\left(
\tilde{\textbf{x}}\right) = 0$, $\left(\forall k\in K\right)~
\mathcal{I}_{k}\left(\tilde{x}\right) \geq 0$ and  $\sum_{k\in K}
\nu_{k}\mathcal{I}_{k} \left(\tilde{\textbf{x}}\right) = 0$, where 
$\tilde{\textbf{x}}$ is the maximum of $\mathcal{L}$ in 
$\mathcal{D}$ for $\left\{\left\{\lambda_{j}\right\}_{J}, \left\{
\nu_{k}\right\}_{K}\right\}$, then $\tilde{\textbf{x}}$ is a 
solution of the original optimization problem. 
\\ \\ 
\emph{Proof.} $\forall \textbf{x}_{*}\in\mathcal{D}_{*}$, 
points satisfying the constraints of the original (\emph{primal}) 
optimization, $\left(\forall k\in K\right)\mathcal{I}_{k}\left(
\textbf{x}_{*}\right) \geq 0$, and $\left(\forall j\in J \right)
\mathcal{E}_{j}\left(\textbf{x}_{*}\right) = 0$. 
Therefore, 
$$m_{*} =\text{max}_{\mathcal{D}_{*}}~\mathcal{O}\leq\text{max}_{
\mathcal{D}_{*}}~\mathcal{L} \leq \text{max}_{\mathcal{D}}~
\mathcal{L}.$$ 
$\text{max}_{\mathcal{D}}~\mathcal{L}$ is convex as it is a sum of 
compositions of convex functions, max and affine functions of $\left
\{\lambda_{j}\right\}_{J}$ and $\left\{\nu_{k}\right\}_{K}$. 
If a collection $\left\{\left\{\lambda_{j}\right\}_{J}, \left\{
\nu_{k}\right\}_{K}\right\}$ is found such that $\left(\forall j\in J
\right)~\mathcal{E}_{j}\left(\tilde{\textbf{x}}\right) = 0$, 
$\left(\forall k\in K\right)~\mathcal{I}_{k}\left(\tilde{x}\right) 
\geq 0$ and $\sum_{k\in K}\nu_{k}\mathcal{I}_{k} \left(\tilde{
\textbf{x}}\right) = 0$ then $\tilde{x}\in\mathcal{D}_{*}$ and 
$$\mathcal{O}\left(\tilde{\textbf{x}}\right) = \mathcal{L}\left(
\tilde{\textbf{x}}\right) = \text{max}_{\mathcal{D}}~\mathcal{L}.
~\square$$ 
\indent Whenever the operator appearing between a field and its 
associated linear functional in a sesquilinear relation is positive 
definite, the constraint describes a compact manifold. 
This is always true of \eqref{asymOpt}, and so, as both constraints 
are closed sets, the domain of \eqref{optProb} is compact. 
Moreover, by the validity of the $\left|\textbf{T}\right> = 
\textbf{0}$ solution, the domain is non-empty. 
As such, \eqref{optProb} is assured to have a unique (non trivial) 
maximum value occurring at some stationary point (or points), and it 
is meaningful to consider the Lagrangian dual
\begin{equation}
\mathcal{G}\left(\zeta,~\gamma\right) = \text{max}_{\mathcal{F}}~\mathcal{L},
\end{equation}
where the domain $\mathcal{F}$ is set by the criterion that 
$\text{max}~\mathcal{L}$ is finite.
Under this assumption, taking partial derivatives over
$\left|\textbf{T}_{\ell}\right>$, a stationary point of 
$\mathcal{L}$ requires $\left(\forall \ell\right)$,
\begin{align}
&\Bigg(\zeta \asym{\mathbb{U}_{\ell}} + 
\gamma \sym{\mathbb{U}_{\ell}} +\mathbb{O}_{\ell} \Bigg)
\left| \textbf{T}_{\ell}\right> =
\nonumber \\
&\frac{i}{2}\left|\textbf{S}^{\left(2\right)}_{\ell}\right> +
\frac{\gamma + i\zeta}{2}\left|
\textbf{S}^{\left(1\right)}_{\ell}\right>.
\label{statConds}
\end{align}
A collection $\left\{\zeta,~\gamma\right\}\in \mathcal{F}$ if and 
only if $\big(\mathbb{O}_{\ell}+\zeta \asym{\mathbb{U}_{\ell}}
+ \gamma \sym{\mathbb{U}_{\ell}}\big)$ is positive definite for all 
$\ell$, and so
\begin{align}
\left(\forall \ell\right)~\mathbb{A}_{\ell} =& 
\mathbb{A}_{\ell}^{\dagger} = \Bigg(\mathbb{O}_{\ell}
+ \zeta \asym{\mathbb{U}_{\ell}}  +
\gamma \sym{\mathbb{U}_{\ell}}  \Bigg)^{-1}
\label{Aopt}
\end{align} 
is inherently both defined and positive definite. 
(If any $\mathbb{A}^{-1}_{\ell}$ is not positive definite, there 
is then a field $\left|\textbf{T}_{\ell}\right>$ such that 
$\mathcal{G}\rightarrow \infty$.) 
Letting $\left|\textbf{S}^{\left(3\right)}_{\ell}\right> = 
\left(\zeta -i\gamma\right)\left|\textbf{S}^{\left(1\right)}_{\ell}
\right> + \left|\textbf{S}^{\left(2\right)}_{\ell}\right>$, and 
$\left|\textbf{S}^{\left(4\right)}_{\ell}\right> = \mathbb{A}_{\ell}
\left|\textbf{S}^{\left(3\right)}_{\ell}\right>$, it follows that 
the unique stationary point of the dual occurs when $\left|
\textbf{T}_{\ell}\right> = \frac{i}{2}\left|\textbf{S}^{\left(4
\right)}_{\ell}\right>$. 
Hence, within $\mathcal{F}$ (using the above replacement),
\begin{equation}
\mathcal{G} = 
\frac{1}{4}\sum_{\ell}\left<
\textbf{S}^{\left(3\right)}_{\ell}\right|
\mathbb{A}_{\ell}\left|
\textbf{S}^{\left(3\right)}_{\ell}\right>.
\label{gSol}
\end{equation}
The gradients of \eqref{gSol} exactly reproduce the constraint
equations $\frac{\partial \mathcal{G}}{\partial \zeta} =
\mathcal{C}_{\zeta}$ and $\frac{\partial \mathcal{G}}{\partial \gamma} = \mathcal{C}_{\gamma}$, with 
\begin{align}
&\mathcal{C}_{\zeta} = 
\sum_{\ell}\frac{1}{2}\re{
\left<\textbf{S}^{\left(1\right)}_{\ell}\Big|
\textbf{S}^{\left(4\right)}_{\ell}\right>} -
\frac{1}{4}\left<\textbf{S}^{\left(4\right)}_{\ell}\right|
\asym{\mathbb{U}_{\ell}}\left|
\textbf{S}^{\left(4\right)}_{\ell}\right>,
\nonumber\\ 
&\mathcal{C}_{\gamma}= \sum_{\ell}\frac{1}{2}\im{\left<
\textbf{S}^{\left(4\right)}_{\ell}\Big|
\textbf{S}^{\left(1\right)}_{\ell}\right>} -
\frac{1}{4}\left<\textbf{S}^{\left(4\right)}_{\ell}\right|
\sym{\mathbb{U}_{\ell}}\left|
\textbf{S}^{\left(4\right)}_{\ell}\right>.
\label{solEqs}
\end{align}
Therefore, if a stationary point within the feasibility region is 
found, strong duality holds. 
The function is convex, and thus has a single stationary point, 
either on the boundary or within the domain of feasibility. 
If the point is within the domain of feasibility, then by 
\eqref{solEqs} the corresponding field satisfies the constraints, 
and by the lemma, the corresponding point is strongly dual. 
In this case, the solution of \eqref{optProb} is
\begin{align}
&\mathcal{O} = \sum_{\ell}\frac{1}{2}
\re{\left<\textbf{S}^{\left(2\right)}_{\ell}\Big|
\textbf{S}^{\left(4\right)}_{\ell}\right>} - 
\frac{1}{4}\left<\textbf{S}^{\left(4\right)}_{\ell}\right|
\mathbb{O}_{\ell}\left|
\textbf{S}^{\left(4\right)}_{\ell}\right>,
\label{objVal}
\end{align}
with $\left\{\zeta,~\gamma\right\}$ set by the simultaneous zero 
point of \eqref{solEqs}. 
If no such point exists in $\mathcal{F}$, then the unique minimum 
value of $\mathcal{G}$ in the domain of feasibility, attained on the 
boundary of some $\mathbb{A}_{\ell}$ becoming semi-definite, remains 
a bound on $\mathcal{O}$ in \eqref{optProb}. 
That is, $\mathcal{O}\leq \sum_{\ell} \left<\textbf{S}^{\left(3
\right)}_{\ell}\right|\mathbb{A}_{\ell}\left|\textbf{S}^{\left(3
\right)}_{\ell}\right>/4$ in all cases. 
Comments on solving \eqref{solEqs} are given in Sec.~\ref{toyBounds}.

\section{Relations to Prior Art}\label{priorWork}
Previous work in the area of electromagnetic performance limits can 
be loosely classified into three overarching strategies: modal 
limits, shape-independent conservations limits, and scattering 
operator limits. 
Each approach presents its own relative strengths and weaknesses. 
Below, we provide a rough sketch of these contributions as they 
relate to this work, particularly the use and interpretation of 
constraints \eqref{symOpt} and \eqref{asymOpt}.

Arguments for response bounds based on modal decompositions,
exploiting quasi-normal, spectral, characteristic, Fourier and/or
multipole expansions~\cite{chu1948physical,yamamoto1986preparation,
mclean1996re,geyi2003physical,hamam2007coupled,kwon2009optimal,
verslegers2010temporal,ruan2010superscattering,ruan2012temporal,
munday2012light,fleury2014physical,liberal2014upper,jia2015theory,
monticone2016invisibility,nordebo2017physical}, have been widely 
considered for many decades. 
And like the classical diffraction and blackbody limits of ray 
optics, they have proven to be of great practical value for 
describing possible interactions between large objects and 
propagating waves~\cite{yu2010grating,hugonin2015fundamental,
miroshnichenko2018ultimate}. 
Yet, in complement, the need to enumerate and characterize what 
modes may possibly participate has also long proved problematic. 
Small sources, separations, and domains typically require many 
elements to be properly represented in any basis well suited to 
analysis of Maxwell's equations, and so, especially in the 
near-field and without knowledge of the geometric characteristics of 
the scattering object, or inclusion of material considerations, 
there is no systematically effective approach to bound
modal sums (without introducing additional aspects as is done in
scattering operator approaches). 
While a variety of considerations (transparency, size, etc.) have 
been heuristically employed in an attempt to introduce reasonable
cut-offs~\cite{miller2000communicating,sohl2007physical,
yu2010fundamental,yu2011angular,yu2012thermodynamic,yang2017low}, 
the values obtained by modal methods in such settings are 
consistently orders of magnitude too large~\cite{
miller2016fundamental,pendry1999radiative,
venkataram2019fundamental}. 
Nevertheless, the idea that modes often separate otherwise muddled 
aspects of photonics remains a key insight. 

Shape-independent conservation limits, utilizing energy~\cite{
callahan2012solar,miller2016fundamental,michon2019limits} 
and/or spectral sum rules~\cite{gordon1963three,fuchs1976sum,
mckellar1982sum,miller2007fundamental,gustafsson2012optical,
miller2014fundamental,cassier2017bounds,shim2019fundamental} to set 
local limits based on physical laws, generally display the opposite 
behavior, and are known to give highly accurate estimates of maximal 
far-field scattering interactions in the limit of vanishingly 
small sizes (quasi-statics) for certain (near resonant) metallic 
materials~\cite{miller2014fundamental,miller2016fundamental,
yang2018maximal,shahpari2018fundamental,capek2019optimal}.
Notwithstanding, as we have found in our work on bounds for radiative
heat transfer~\cite{venkataram2019fundamental,molesky2020fundamental}
and angle-integrated radiative emission~\cite{molesky2019bounds}, 
they are not sufficient in and of themselves to properly capture 
various relevant, performance limiting, wave effects. 
Intuitively,  without any geometric information, a conservation 
argument must apply on a per volume basis, which is at odds with the 
area scaling of ray optics.

As a relevant example, consider the fact that the global power
quantities given in Sec.~\ref{formalism} must be non-negative. 
Two of these turn out to be unique and thus set physically motivated 
algebraic constraints on the $\mathbb{T}$ operator. 
For any vector field $\left|\textbf{E}\right>$, the non-negativity 
of scattering (known as passivity~\cite{liu2013causality}) imposes 
that $$\left<\textbf{E}\right|\asym{\mathbb{T}} - \mathbb{T}
^{\dagger}\asym{\mathbb{V}^{-1\dagger}}\mathbb{T}\left|\textbf{E}
\right> \geq 0,$$ while the non-negativity of absorption imposes
that $$\left<\textbf{E}\right|\asym{\mathbb{T}} - \mathbb{T}
^{\dagger}\asym{\mathbb{G}^{\text{0}}}\mathbb{T}\left|\textbf{E}
\right> \geq 0.$$ Both conditions are concurrently captured in
\eqref{asymOpt}, which is equivalent to the optical 
theorem~\cite{jackson1999classical}: physically, the sum of the 
absorbed power and scattered power, \eqref{fAbs} and \eqref{fSct}
$\mathbb{T}^{\dagger}_{ss}\left(\asym{\mathbb{G}^{\text{0}\dagger} +
  \mathbb{V}^{-1\dagger}}\right)\mathbb{T}_{ss}$, must equal the
extracted power, \eqref{fExt} $\asym{\mathbb{T}_{ss}}$.  If no
additional information pertaining to possible geometries or the
characteristics of $\left|\textbf{E}\right>$ is given, then the most
that can be said from \eqref{asymOpt} is that no singular value of
$\mathbb{T}$ can be larger than the inverse of the smallest singular
value of $\asym{\mathbb{V}^{-1\dagger}}$.  For a single material with
response characterized by a local electric susceptibility $\chi$, this
logic yields the material loss figure of merit
\begin{equation}
\lVert\mathbb{T}\rVert \leq \zeta_{\text{mat}} =
\frac{\left|\chi\right|^{2}}{\im{\chi}},
\label{zetaMat}
\end{equation} 
originally derived in Ref.~\cite{miller2016fundamental} directly 
using the implications of passivity for polarization fields. 
The universal applicability of this largest possible response has 
profound consequences for the design of many photonic devices 
relying on weakly metallic response ($-1 \gtrsim \re{\chi} \gtrsim 
-10$) and small interaction volumes~\cite{yang2017low,
yang2018maximal}. 
In these cases, it is often fair to assume that a resonance can be 
created and that $\asym{\mathbb{G}^{\text{0}}}\approx \textbf{0}$. 
Little structuring is required to achieve a plasmonic resonance, and 
the maximum achievable polarization current is indeed dominated by 
material losses. 
But, for single material devices where light-matter interactions 
occur on length scales comparable to (or greater than) the 
wavelength ($\gtrsim \lambda/10$), or the real part of $\chi$ is
outside the range stated above (e.g., strong metals or dielectrics), 
such estimates are overly optimistic for single material devices, 
Sec.~\ref{applications}. 
Over a large enough domain, the generation of polarization currents 
capable of interacting with propagating fields leads to radiative 
losses which have been neglected by supposing $\asym{\mathbb{G}
^{\text{0}}}\approx\textbf{0}$. 
To create a bright (active) far-field resonance, it must be possible 
to couple to radiation modes and then confine the resulting 
generated field within the domain. 
This is not always possible for any prescribed choice of material 
and device size.

Scattering operator approaches aim to eliminate the weaknesses of 
modal and shape-independent conservation arguments by combining 
their strengths~\cite{baum1986bounds,miller2000communicating,
gustafsson2007physical,tsang2011fundamental,liberal2014least,
miller2015shape,ang2017quantum,miller2017universal,
zhang2019scattering,molesky2019bounds,molesky2020fundamental}.
Innately, the Green function of an encompassing domain (through 
its link to Maxwell's equations) provides both a modal basis for, 
and constraints on, modal sums. 
In concert, restrictions on the possible characteristics 
of the $\mathbb{T}$ operator can be used to ensure that physical 
laws and scaling behavior are properly observed. 
A number of encouraging conclusions have been derived in this 
manner. 
Drawing from our own work, in Ref.~\cite{molesky2020fundamental} 
it was shown that imposing \eqref{zetaMat} on the operator 
expression for angle-integrated absorption and thermal emission, 
\eqref{themralEmission}, is sufficient to generate bounds smoothly 
transitioning from the absorption cross section limit of a resonant 
metallic nanoparticles (the product of the volume and 
$\zeta_{\text{mat}}$) to the macroscopic blackbody limits of ray 
optics. 
Similar methods were used in Ref.~\cite{venkataram2019fundamental} 
to prove that, for equal values of $\zeta_{\text{mat}}$, 
nanostructuring cannot appreciably improve near-field thermal 
radiative heat transfer compared to a resonant planar system. 

Nevertheless, careful investigation of the previous situations where 
scattering operator amalgamations have been successfully applied 
reveals a consistent use of niceness properties that are not 
generally valid. 
In the examples given above, we were aided by the fact that thermal 
sources are completely uncorrelated and, for thermal emission and 
integrated absorption, that only propagating fields needed to be 
treated. 
Without these helpful aspects, there are situations where past
scattering operator approaches, which focused exclusively on
conservation of real power \eqref{asymOpt}, would add complexity
without tightening the asymptotics provided by shape-independent
conservation arguments (dashed lines in the figures of
Sec.~\ref{applications}). 
Furthermore, using the currently practiced technique of translating 
established physical principles back to implied operator properties 
and then using inequality compositions to produce limits, it is 
difficult to see how the interaction of more than one or two 
additional constraints could ever possibly be accounted for. 
The switch to exploiting algebraic deductions beginning from 
\eqref{Tformal} in combination with standard optimization theory 
may seem a subtle distinction; however, the flexibility offered by 
Lagrange duality suggests that this view may be of substantial 
benefit going forward. 

A related shift towards systemization has been realized in
the recent report of computational bounds by Angeris, 
Vu{\v{c}}kovi{\'c}, and Boyd~\cite{angeris2019computational}, 
treating linear electromagnetics as an optimization problem with 
respect to a target field (or a collection of target fields). 
The result, also making of use Lagrange duality, has immediate 
consequences for qualitatively understanding and improving inverse 
design. 
Yet, it does not allow one to make conclusive statements about 
feasibility and relative merits, as is true of traditional 
limits. 
More precisely, what is found is a ``computational certificate'': 
given a target field and an evaluation metric, the algorithm returns 
a number; any vector satisfying Maxwell's equation will have a 
metric disagreement with the target at least as large as the number. 
That is, the algorithm does not find physical limits, but instead a 
minimum bound on distance, in a certain user determined measure, 
between a particular field and the set of physically possible 
fields. 
There may be situations where this difference is of little 
consequence, or provably zero, but a priori there are no guarantees. 
There need not be any relation between the value taken by a function 
at a point and how near that point is to some set. 

Finally, while concluding the writeup and peer review of this article,
we have become aware of contemporary works by Gustafsson et
al.~\cite{gustafsson2019upper}, extending developed methods for
bounding the performance of radio frequency antennas~
\cite{gustafsson2012physical,gustafsson2016antenna,
shahpari2018fundamental,capek2019optimal}, Kuang \textit{et al.}~
\cite{kuang2020maximal}, and Trivedi \textit{et al.}~
\cite{trivedi2020fundamental}.  Independently and simultaneously
developed, the first two formulations are in many respects quite like
the method presented here.  Working from the perspective of
polarization currents, both articles bound the optimization of
objectives equivalent to \eqref{fAbs}--\eqref{cMat} subject to power
constraints via Lagrangian duality.  Ref.~\cite{gustafsson2019upper}
incorporates both \eqref{symOpt} and \eqref{asymOpt} while
Ref.~\cite{kuang2020maximal} uses only \eqref{asymOpt}, leading to 
the discrepancies between the dashed and solid lines depicted in
Sec.~\ref{applications}.  Taken together the two articles present a
rich collection of findings reinforcing some of the observation of
Sec.~\ref{applications}.  Nonetheless, tangible differences do exist.
Although largely unexploited in this initial investigation, there are
many advantages offered by working with operator as opposed to vector
relations for further generalizations.  In contrast, the more recent
work of Trivedi \textit{et al.}~\cite{trivedi2020fundamental} focuses on how
the physically required self-consistency of scattering theory sets
constraint on the possible properties of the scattered electric 
field.
Branching off from this alternative starting point, the optimization
problem and subsequent application of Lagrange duality acquires
characteristics quite different from those encountered here. 
The closest comparison would appear to be something akin to imposing 
only the conservation of reactive power, as the approach described in
Ref.~\cite{trivedi2020fundamental} appears to produce non-trivial 
bounds only when resonant response is not possible.

\section{Computational Mechanics and Single Channel Asymptotics}
\label{toyBounds}

To elucidate the mechanics of \eqref{optProb}, we now describe the
basics of our computational procedure for the specific example of a
spherical confining boundary. 
The discussion is broken into two subsections. 
The first sketches an outline of the approach by which the results 
of Sec.~\ref{applications} for compact bodies were obtained. 
The second considers a simplified single-channel (family) version of
\eqref{optProb} that becomes exact as $R\rightarrow 0$, predicting
many of the trends seen in Fig.~2 and Fig.~3 of Sec.~
\ref{applications}. 
Namely, the largest possible interaction enhancements are found to 
obey either an effective medium ``dilution'' response, or the 
material dependence encountered in Rayleigh scattering~\cite{
hulst1981light}. 
For low-loss dielectrics ($\re{\chi} > 0$) and strong metals
($\re{\chi}\ll -1$), this can lead to large discrepancies with 
respect to previously established per volume bounds based on the 
material loss figure of merit $\zeta_{\text{mat}}$~\cite{
miller2016fundamental,yang2017low,yang2018maximal,michon2019limits,
molesky2019bounds,molesky2020fundamental}.

\subsection{Computational Mechanics}

Recall that, once an origin has been specified, the Green function 
can always be expanded in terms of the regular (finite), 
$\textbf{RN}$ and $\textbf{RM}$, and outgoing, $\textbf{N}$ and 
$\textbf{M}$, spherical wave solutions to Maxwell's equations 
as~\cite{tsang2004scattering,kruger2012trace}
\begin{align}
&\mathbb{G}^{\text{0}}\left(\textbf{x},\textbf{y}\right) = -
\int\limits_{\text{Y}}\delta\left(\textbf{x}-\textbf{y}\right) 
\hat{x}\otimes\hat{y} + i 
\sum_{\ell = 1}^{\infty}
\sum_{m=-\ell}^{\ell} \left(-1\right)^{m}\int\limits_{\text{Y}}
\nonumber \\
&\begin{cases}
\textbf{M}_{1,m}\left(\textbf{x}\right)
\textbf{RM}_{\ell,-m}\left(\textbf{y}\right) 
+ \textbf{N}_{\ell,m}\left(\textbf{x}\right)
\textbf{RN}_{\ell,-m}\left(\textbf{y}\right),~x > y\\
\textbf{RM}_{\ell,m}\left(\textbf{x}\right)
\textbf{M}_{\ell,-m}\left(\textbf{y}\right) 
+ \textbf{RN}_{\ell,m}\left(\textbf{x}\right)
\textbf{N}_{\ell,-m}\left(\textbf{y}\right),~x < y.
\end{cases}
\label{GreenExpand}
\end{align}
In \eqref{GreenExpand}, $\textbf{x}$ and $\textbf{y}$ are used to
denote the wave vector normalized radial vectors of the domain and
codomain, i.e., $\textbf{x} = \left<2\pi r/\lambda,\theta,
\phi\right>$, with $x$ and $y$ used for the corresponding radial 
parts. 
The integral over $\text{Y}$ is taken to mean integration over the 
$\textbf{y}$ coordinate. 
Note that there is no complex conjugation in these integrals, and 
that our notation for the Green function is unconventional in that 
an additional factor of $k^{2} = \left(2\pi/\lambda\right)^{2}$ 
is included as part of the definition, as opposed to a separate 
conversion factor, allowing all spatial distances to be normalized 
in terms of the wavelength. 
So long as the current source is not located within the domain in
question, any incident field can be expanded in term of the
regular waves~\cite{wittmann1988spherical,tsang2004scattering,
kruger2012trace}. 
As such, the spectral basis of the asymmetric part of 
\eqref{GreenExpand}
\begin{align}
\asym{\mathbb{G}^{\text{0}}} = 
\sum_{\ell,m}\left(-1\right)^{m} &\int
\limits_{\text{Y}}\textbf{RM}_{\ell,m}\left(\textbf{x}\right)
\textbf{RM}_{\ell,-m}\left(\textbf{y}\right) + 
\nonumber \\
&\textbf{RN}_{\ell,m}\left(\textbf{x}\right)
\textbf{RN}_{\ell,-m}\left(\textbf{y}\right),
\end{align}
the unit normalized $\hat{\textbf{RM}}_{\ell,m}$ and $\hat{
\textbf{RN}}_{\ell,m}$ waves, serves as convenient choice for 
starting point for generating the basis vector families 
required to properly represent $\mathbb{U}_{\ell}$.\footnote{As 
confirmed by the examples explored here, $\sym{\mathbb{G}^{\text{0}}}$ 
and $\asym{\mathbb{G}^{\text{0}}}$ are usually not simultaneously 
diagonalizable~\cite{harrington1972characteristic}.}
That is, given the form of the regular solutions 
\begin{align}
&\textbf{RN}_{\ell,m}\left(\textbf{y}\right) = 
\frac{\sqrt{\ell+1}}{y}\text{j}_{\ell}\left(y\right)
\text{A}^{\left(3\right)}_{\ell,m} + 
\frac{1}{y}
\frac{\partial\left(y~\text{j}_{\ell}\left(y\right)\right)}{\partial y}
\text{A}^{\left(2\right)}_{\ell,m},
\nonumber \\
&\textbf{RM}_{\ell,m}\left(\textbf{y}\right) = ~\text{j}_{\ell}\left(y\right)
\text{A}^{\left(1\right)}_{\ell,m},
\label{regSols}
\end{align}
the orthonormality of the vector spherical harmonics ($\text{A}^{
\left(1\right)}_{\ell,m}$, $\text{A}^{\left(2\right)}_{\ell,m}$, and 
$\text{A}^{\left(3\right)}_{\ell,m}$, see
Ref.~\cite{kristensson2014spherical} for details) means that the
Green function \eqref{GreenExpand} does not couple the $\ell, m$ or 
$\textbf{RN}$ and $\textbf{RM}$ labels. 
Hence, the individual radiation channels act as an effective 
partitioning, and by taking these vectors as the ``seeds'' or 
``family heads,'' a complete (simplifying) basis for \eqref{optProb} 
can be generated through the Arnoldi (Krylov subspace) 
procedure~\cite{stoer2013introduction}. 
Briefly, starting with a given unit normalized regular wave, $\hat{
\textbf{RN}}_{\ell,m}$, one generates $\mathbb{U}\left|\hat{
\textbf{RN}}_{\ell,m}\right> = \left(\mathbb{V}^{\dagger -1} - 
\mathbb{G}^{\text{0}\dagger}\right)\left|\hat{\textbf{RN}}_{\ell,m}
\right>$. 
Projecting out the $\hat{\textbf{RN}}_{\ell,m}$ component of this 
image and normalizing, one obtains a new vector
$\left|\hat{\textbf{PN}}_{\ell,m}^{\left(2\right)}\right>$. 
$\left|\hat{\textbf{PN}}_{\ell,m}^{\left(2\right)}\right>$ 
then serves as the input for the next iteration, and in this way the 
$\ell$ family (block), more properly the $\hat{\textbf{RN}}
_{\ell,m}$ family, of the matrix representation of the $\mathbb{U}$ 
operator ($\mathbb{U}_{\ell}$) is computed, i.e., $\mathbb{U}_{\ell}$ 
is represented in the basis $\left\{\left|\hat{\textbf{RN}}_{\ell,m}
\right>,~\left|\hat{\textbf{PN}}_{\ell,m}^{\left(2\right)}\right>,
~\left|\hat{\textbf{PN}}_{\ell,m}^{\left(3\right)}\right>,~
\ldots\right\}$.

Technically the above process does not terminate, but regardless,
two practical consideration lead to workable numerical
characteristics.\footnote{Care is needed to avoid numerical 
instability, see \ref{appendix} B. for details.} 
First, due to the fact that each vector is orthogonal to all others, 
the off-diagonal coupling components of $\mathbb{U}_{\ell}$ in every 
family originate entirely due to the volume integrals in 
\eqref{GreenExpand}. 
Therefore, in the limit of vanishing volume, or high $\ell$, each 
$\mathbb{U}_{\ell}$ is effectively $2\times 2$. 
Second, by the Arnoldi construction, all upper diagonals beyond 
$\text{diag}_{1}$, with $\text{diag}_{0}$ standing for the main, are 
zero. 
Thus, because at each step the generation of new basis components is 
driven entirely by $\sym{\mathbb{G}^{\text{0}}}$, and $\sym{
\mathbb{G}^{\text{0}}} = \sym{\mathbb{G}^{\text{0}}}^{\dagger}$, 
the matrix representation of every $\mathbb{U}_{\ell}$ is 
tridiagonal. 
Due to this banded nature, each $\mathbb{A}_{\ell}^{-1}$ 
gives a simple, conclusive, estimate of the error for images 
generated by $\mathbb{A}_{\ell}$. 
All that is required is to pad the current solution with zeros and 
calculate its image under $\mathbb{A}_{\ell}^{-1}$ in a basis 
augmented by three additional elements. 
The magnitude of the error of the image compared to the source is 
exactly the same as would be found in \emph{any} larger (even 
infinite) basis, see Sec.~\ref{appendix} for additional details. 

Coupled with \eqref{gSol}, the determination of bounds for any 
electromagnetic process in a compact body (an object of bounded 
extent) that can be described as a total absorption, scattering or 
extinction process, is thus mapped to the numerical determination of 
the minima of a constrained convex function. 
Many efficient algorithms exist to solve such problems~\cite{
moreau2003coordinated,ozban2004some,johnson2014nlopt}, along with a 
variety of nice introductions~\cite{boyd2004convex,
stoer2013introduction,nesterov2018lectures}.

\subsection{Single Channel Asymptotics}

Consider the simplified optimization problem
\begin{align}
&\text{max}~
\left<\textbf{T}_{\ell}^{\left(1\right)}\right|\mathbb{P}_{1}
\left|
\textbf{T}_{\ell}^{\left(1\right)}\right>~\text{such that}
\nonumber \\
&\mathcal{C}_{\zeta} = \im{\left<
\textbf{S}^{\left(1\right)}_{\ell}\Big|
\textbf{T}_{\ell}\right>} - \left<\textbf{T}_{\ell}
\Big|\asym{\mathbb{U}_{\ell}}
\Big|\textbf{T}_{\ell}\right> = 0,
\nonumber \\
&\mathcal{C}_{\gamma} = 
\re{\left<\textbf{S}^{\left(1\right)}_{\ell}
\Big|\textbf{T}_{\ell}\right>} - 
\left<\textbf{T}_{\ell}\Big|\sym{\mathbb{U}_{\ell}}
\Big|\textbf{T}_{\ell}\right> = 0,
\label{limitOpt}
\end{align}
where it has been assumed that that $\left(\forall f_{i}\neq\ell
\right) ~\left<\textbf{S}_{f_{i}}\right| = \textbf{0}$,  and $
\mathbb{P}_{1}$ represents the projection of $\left|\textbf{T}\right>
$ onto the $\ell = 1$ ``family head.'' 
So stated, \eqref{limitOpt} represents the maximum possible 
interaction that can occur between a generated polarization current 
and an exciting field for a single radiation channel, respecting the 
conservation of total power. 
Based on the Taylor series representation of the spherical Bessel 
functions,
$$\text{j}_{\ell}\left(kR\right) = \sum_{q=0}^{\infty}
\frac{\left(-1\right)^{q}}{q!\left(2\ell+2q+1\right)!!}
\left(\frac{kR}{2}\right)^{q},$$ 
there are two situations in which this problem is fairly simple to 
treat analytically. 
If either the radius $kR \ll 1$, with $k = 2\pi/\lambda$, or 
$\ell$ is large compared to $kR$, then both the regular and 
outgoing waves appearing in the Green's function, 
\eqref{GreenExpand}, are well approximated by two-term expansions. 
This feature causes the above Arnoldi procedure for basis generation 
to effectively terminate after constructing a single image vector. 
As high $\ell$ contributions (for propagating waves) occur 
only when large low $\ell$ contributions are also present, no 
meaningful asymptotic behavior can be extracted from a single 
channel analysis of the second possibility. 
Accordingly, we will focus on the assumption that $kR \ll 1$.  
Symbolically carrying out the required image generation and 
orthogonalization steps, the representation of $\mathbb{U}_{f_{1}}$ 
in this quasi-static ($R\rightarrow 0$) regime is
\begin{align}
\mathbb{U}_{f_{1}} &= \mathbb{V}^{\dagger -1} - 
  \mathbb{G}^{\text{0}\dagger}_{f_{1}} 
  =\sym{\mathbb{U}_{f_{1}}} + 
  i\asym{\mathbb{U}_{f_{1}}} 
  \nonumber \\
  &=\begin{bmatrix}
  \frac{1}{3}+\frac{\re{\chi}}{\left|\chi\right|^{2}}-\frac{4\left(kR\right)^{2}}{15} & -\frac{\left(kR\right)^{2}}{5\sqrt{14}} \\
  -\frac{\left(kR\right)^{2}}{5\sqrt{14}}& 
  \frac{\re{\chi}}{\left|\chi\right|^{2}}-\frac{2\left(kR\right)^{2}}{45}
  \end{bmatrix}
  \nonumber \\
  &\hspace{1in}+i
  \begin{bmatrix}
  \frac{\im{\chi}}{\left|\chi\right|^{2}}+
  \frac{2\left(kR\right)^{3}}{9} & 
  0 \\
  0 & 
  \frac{\im{\chi}}{\left|\chi\right|^{2}}
  \end{bmatrix}.
  \label{qsMat}
  \end{align}
Within its regime of validity, \eqref{qsMat} has two key features. 
First, due to the identity portion of the Green's function, the 
$\left(1,1\right)$ element has a constant positive piece in addition 
to the $\re{\chi}/\left|\chi\right|^{2}$ contribution made by
$\mathbb{V}^{\dagger -1}$. 
Second, all off-diagonal elements are small. 
The first feature sets a critical material response value for which 
it is possible that the $\left(1,1\right)$ element of 
$\sym{\mathbb{U}_{1}}$ may be negative: $\re{\chi} \leq -3$. 
The second feature allows off-diagonal terms to be neglected in 
comparison to diagonal terms for most ($\left|\re{\chi}\right|$ 
sufficiently large) susceptibility values. 

Denoting the symmetric and anti symmetric components of the 
representation of $\mathbb{U}_{1}$ as $u^{\left(\cdot,\cdot
\right)}_{s}$ and $u^{\left(\cdot,\cdot\right)}_{a}$, solving 
\eqref{optProb} amounts to determining the $\left\{t_{1},t_{2}\right
\}$ component pair producing the largest magnitude $t_{1}$ such that
\begin{align}
&\text{sin}\left(\theta\right)s_{1}t_{1} - 
t_{1}^{2}u_{a}^{\left(1,1\right)} - t_{2}^{2}
u_{a}^{\left(2,2\right)} = 0, 
\nonumber \\
&\text{cos}\left(\theta\right)s_{1}t_{1} - 
t_{1}^{2}u_{s}^{\left(1,1\right)} - t_{2}^{2}
u_{s}^{\left(2,2\right)} 
+ 2\text{cos}\left(\phi\right) t_{1}t_{2}u^{\left(1,2\right)}_{s} =
0.
\label{limProbAtomic}
\end{align}
Here, the $t_{1}$ and $t_{2}$ variables are the (positive) magnitude
coefficients of $\left|\textbf{T}_{1}\right>$ in the first and second
Arnoldi vectors of the $\ell = 1$ family, $s_{1}$ is the coefficient 
of the source, $\theta$ is the relative phase difference between the 
source and first coefficient of $\left|\textbf{T}_{1}\right>$, and 
$\phi$ is the relative phase difference within the two coefficients 
of $\left|\textbf{T}_{1}\right>$. 
As a response operator, $\asym{\mathbb{T}}$ must be positive 
semi-definite and so 
$\theta\in\left[0,\pi\right]$~\cite{landau2013statistical}. 
Using the symmetric constraint to solve for $t_{2}$ in terms of 
$t_{1}$, forgetting off-diagonal terms when they appear as sums 
against diagonal terms in the resulting quadratic equation, the 
asymmetric constraint determines that the maximum polarization 
current that can couple to the source is 
\begin{equation}
\zeta_{\text{eff}} =
\frac{t_{1}}{s_{1}} = \frac{\text{cos}\left(\theta\right)
u_{a}^{\left(2,2\right)} -\text{sin}\left(\theta\right)
u_{s}^{\left(2,2\right)}}{u_{a}^{\left(2,2\right)}
u_{s}^{\left(1,1\right)} - u_{a}^{\left(1,1\right)}u_{s}^{\left(2,2\right)}},
\label{ratioVal}
\end{equation} 
subject to the condition, resulting from the requirement that $t_{1}$
and $t_{2}$ are real, that 
\begin{align}
&\left(\text{sin}\left(\theta\right)u_{s}^{\left(1,1\right)}-
\text{cos}\left(\theta\right)u_{a}^{\left(1,1\right)}\right)
\times 
\nonumber \\
&\left(\text{cos}\left(\theta\right)u_{a}^{\left(2,2\right)}-
\text{sin}\left(\theta\right)u_{s}^{\left(2,2\right)}\right)\geq 0.  
\nonumber
\end{align}
Neglecting all higher order corrections but the
radiative efficacy $\rho^{_{\mathbb{G}\textbf{N}}}_{1} =
\asym{\mathbb{G}^{\text{0}}}^{\left(1,1\right)}$, the 
relative magnitude $t_{1}/s_{1} = \zeta_{\text{eff}}$ is therefore 
limited by
\begin{align}
\zeta_{\text{eff}}\leq 
\begin{cases}
\frac{\left|\chi\right|}{\left|\re{\chi}\right|}
\frac{1}{\rho^{_{\mathbb{G}\textbf{N}}}_{1}
+\delta^{_{\mathbb{G}\textbf{N}}}_{1}\im{\chi}/\left|\re{\chi}\right|}
&\frac{\left|\re{\chi}\right|}{\left|\chi\right|^{2}}\leq 
\delta^{_{\mathbb{G}\textbf{N}}}_{1},
\vspace{0.1in}\\
\frac{1}
{\sqrt{\left(\delta^{_{\mathbb{G}\textbf{N}}}_{1} + 
\re{\chi}/\left|\chi\right|^{2}\right)^{2} + 
\left(\rho_{\mathbb{G}} + \im{\chi}/\left|\chi\right|^{2}\right)^{2}}}
&\re{\chi}\geq 0
\end{cases},
\label{limResp}
\end{align}
where $\delta^{_{\mathbb{G}\textbf{N}}}_{1}$ is the domain 
dependent, material independent, portion of $u_{s}^{\left(1,1\right)}
-u_{s}^{\left(2,2\right)}$~\cite{chen1977simple,
yaghjian1980electric}. 
In \eqref{limResp}, the selected symbols are motivated by three 
factors: (1) analogs to the analysis given above 
likely exist whenever all dimension of the design domain are small 
compared to the wavelength, (2) 
$\delta^{_{\mathbb{G}\textbf{N}}}_{1}$ is the delta function portion 
of \eqref{GreenExpand} for the regular $\textbf{RN}_{1,m}$ wave of a 
spherically bounded domain, 
$\delta^{_{\mathbb{G}\textbf{N}}} = 1/3$, and (3) we have 
previously used $\rho^{_{\mathbb{G}\textbf{N}}}_{1} = \frac{2}{9} 
\left(\frac{2\pi R}{\lambda}\right)^{3}$ to denote ``radiative 
efficacy'' in studying the existence of bounds on radiative thermal 
power transfer~\cite{molesky2020fundamental,
venkataram2019fundamental}.  
Utilizing $\left|\textbf{T}\right> = \zeta_{\text{eff}}\left|
\textbf{S}_{1}\right>$ in the power forms provided in 
Sec.~\ref{formalism} produces the single channel cross section 
limits seen in Sec.~\ref{applications}, \eqref{nonResSmallR} and 
\eqref{resSmallR}. 

Full solutions of \eqref{optProb} for an incident planewave are found
to be accurately predicted by \eqref{limResp} as $R\rightarrow 0$,
Fig.~2 and Fig.~3, outside the $-3\leq \re{\chi}\leq -1$ region where
the assumption that the $u_{s}^{\left(1,2\right)}$ terms can be
neglected does not hold.  While the monotonicity property of these
bounds means that (properly scaled) they are accurate for any compact
domain geometry, and one may reasonably guess that the 
characteristics of the Arnoldi process on which the above arguments 
rest are similar in any small volume limit, it should be kept in 
mind that other domain geometries (e.g. 
ellipsoids~\cite{miller2014fundamental,miller2016fundamental}) may 
well display stronger \emph{per volume} response, if the resonance 
condition shifts to more negative values of $\re{\chi}$ 
($\delta^{_{\mathbb{G}\textbf{N}}}_{1} > \nicefrac{1}{3}$). 
Still, while the generated polarization current and exciting 
planewave may exhibit a larger average interaction within the volume 
of a structured body, the net enhancement will be weaker than 
$\zeta_{\text{eff}}$ once the ratio of its volume to an encompassing 
ball is accounted for. 

Given the revealing nature of this simple symbolic analysis for
spherically bounded domains, it is reasonable to ask why we have not
included corresponding results for periodic films. 
Briefly, the reason for this omission has to do with the altered 
proprieties of $\mathbb{G}^{\text{0}}$ that occur in the presence of 
two periodic boundary conditions. 
Unlike the case of the compact domain, in the limit of vanishing 
film thickness $t\rightarrow 0$, important $\delta$ function 
contributions of $\mathbb{G}^{\text{0}}$ are simultaneously
spread across many in-plane wave vectors, with no one channel
containing all key characteristics. 
The only case where single channel asymptotics reproduce observed 
behavior is for extremely thin (non resonant) dielectric films. 

\section{Applications}\label{applications}

In this section, the program developed above is exemplified for two
canonical scattering processes: limits on absorbed and scattered power
for a planewave incident on any structure that could be confined
within a ball of radius $R$, characterizing the maximum cross section
enhancement a body may exhibit~\cite{tsang2004scattering} subject to
the constraints that net real and reactive power are conserved; and
limits on the absorptivity of a periodic film for a normally incident
plane wave as function of the film thickness, $t$.  Because of the
complete exploration of structural possibilities conducted in
calculating these bounds, and the domain monotonicity property
explained in Sec.~\ref{formalism}, all results are equally applicable
to any subdomain, or disconnected collection of subdomains, that fit
inside any given ball or periodic film; therefore, in addition to any
possible connected structures, the bounds pertain to geometries like
arrays of Mie scatterers~\cite{kruk2017functional,
  zhang2019scattering} or plasmonic building
blocks~\cite{zhou2013lasing,tseng2017two}.  Throughout the section,
$R$ and $t$ are unnormalized unless otherwise stated, and $\ell$
stands for the angular momentum number (spherical harmonics
$\text{A}^{\left(-\right)}_{\ell}$), originating from representing
\eqref{optProb} using the Arnoldi construction described in
Sec.~\ref{toyBounds}.

\begin{figure*}[t!]
\centering
\includegraphics[width=2.0\columnwidth]{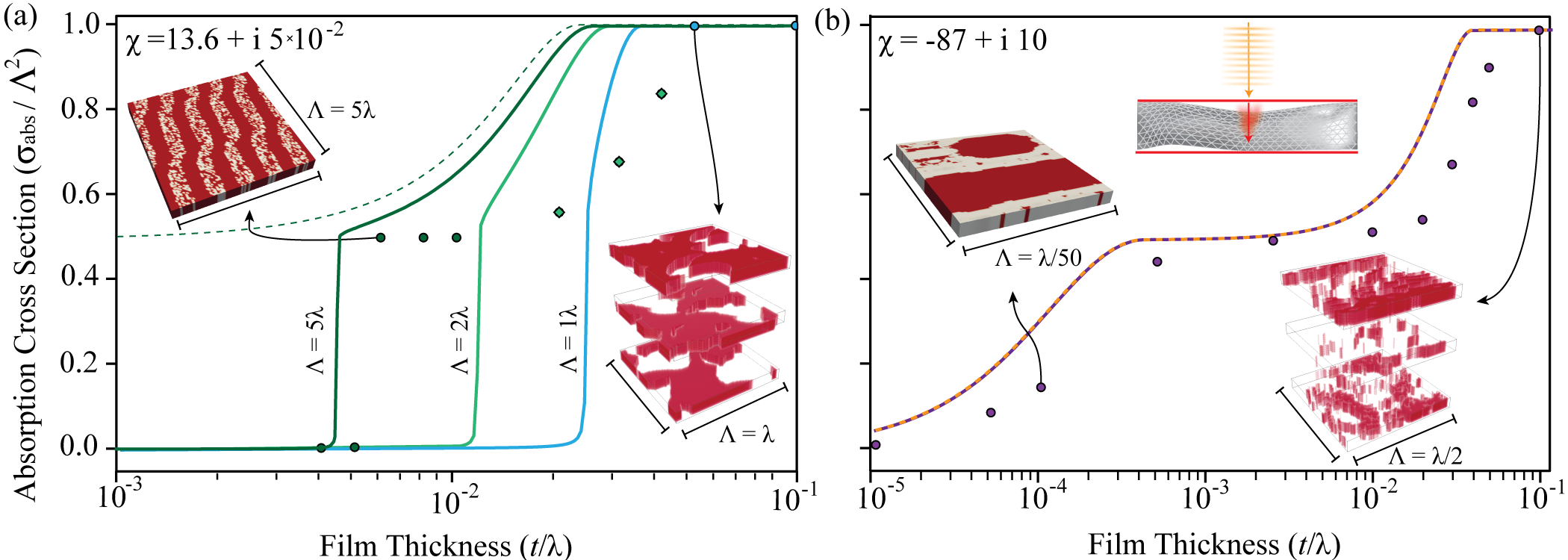}
\vspace{-10 pt}
\caption{\textbf{Absorption cross section bounds for a planewave 
incident on a periodic film.} 
Absorption cross section bounds for any periodic structure 
(e.g. grating, metasurface) of period $\Lambda$ and 
electric susceptibility (a) $\chi = 13.6 + i~5\cdot 10^{-2}$ 
(silicon at $0.8\mu m$) and (b) $\chi = -87 +i~10$ (gold at $1.5\mu m$) 
that can be contained within a planar boundary of 
thickness $t/\lambda$. 
Dots correspond to actual absorption cross section values 
discovered using inverse design following the algorithm described in 
Ref.~\cite{jin2020inverse}. 
A number of visualizations of the associated structures, along with 
a schematic of the absorption process investigated, are included 
as insets. 
All discovered designs with $t/\lambda < 10^{-2}$ consist of a 
single layer, uniform in the normal direction. 
For larger thickness, optimization of three equally thick uniform 
layers is needed to achieve the reported absorption cross sections 
(eventually reaching absorption saturation).
Like prior figures, solid lines are found by solving 
\eqref{optProb} under the constraint that total power is conserved, 
while dashed lines, coinciding with the solid line in (b), are 
found by only imposing the constraint of real power conservation. 
As shown in (a), total power conservation bounds for dielectric 
media depend on the square period supposed. 
Moving from left to right, the periods used in the inverse design 
optimizations are 
$\Lambda = \left\{5\lambda,~5\lambda,~5\lambda,~5\lambda,
~4\lambda,~2\lambda,~3\lambda/2,~3\lambda/2,~\lambda,
~\lambda\right\}$. 
Results for metals ($\re{\chi} < 0$) do not depend on 
the periodicity of the system. 
Discussion of additional bound features relating solely to 
imposing the conservation of real can be found in 
Ref.~\cite{molesky2019bounds} and Ref.~\cite{kuang2020maximal}.}
\vspace{-10 pt}
\end{figure*}

For both spherically bounded examples, the distribution of the power 
density within the domain between different angular momentum numbers 
is strongly tethered to the radius of the boundary. 
Specifically, the coefficients of the electric field, for a unit 
normalized electric field amplitude, in terms of the regular (finite 
at the origin) $\textbf{RM}_{\ell,m}$ and $\textbf{RN}_{\ell,m}$ 
solutions of Maxwell's equations in spherical coordinates, 
\eqref{regSols}, are
\begin{align}
\unitv{E}_{i} = \sum_{\ell=1}^\infty\sum_{\pm }
&i^{\ell+1} \sqrt{(2\ell+1)\pi}~\textbf{RM}_{\ell,\pm 1}
\left(r,\theta,\phi\right) \pm
\nonumber \\
&i^{\ell + 1} 
\sqrt{\left(2\ell+1\right)\pi}~\textbf{RN}_{\ell,\pm 1}
\left(r,\theta,\phi\right),
\label{planeWave}
\end{align}
with $r$ standing for the wave vector normalized radius (the product 
of the true radius and $k = 2\pi/\lambda$). 
Through $\asym{\mathbb{U}}$ \eqref{planeWave} establishes a link 
between the magnitude of the radiative efficacy of each 
channel~\cite{molesky2020fundamental}, or $\ell$ number, and its 
potential for enhancing scattering cross sections: if the planewave 
expansion coefficient of a given channel is large, then so is 
$\asym{\mathbb{G}^{\text{0}}_{\ell}}$, tightening the constraint 
that real power must be conserved.
This leads to a complementary action of the power constraints. 
For any given combination of material and radius, save $\re{\chi} = 
-3$ in the $R\rightarrow 0$ limit, either real or reactive power 
conservation limits induced polarization currents in the medium more 
severely than what would be expected based solely on the material 
loss figure of merit
\begin{equation}
\zeta_{\text{mat}} = 
\frac{\left|\chi\right|^{2}}{\im{\chi}},
\label{matIntFig}
\end{equation}
widely considered in past work on electromagnetic bounds for 
arbitrary materials and structures~\cite{miller2016fundamental,
yang2017low,yang2018maximal,molesky2020fundamental}. 
(An explanation of the origin and usefulness of this quantity is 
given in Sec.~\ref{priorWork}.) 
In Fig.~2 and Fig.~3, dashed curves depict cross section limits 
attained when only the conservation of real power, \eqref{asymOpt}, 
is imposed, as in Ref.~\cite{kuang2020maximal}, and solid curves 
result when reactive power conservation is additionally included. 
All results are found using the Lagrange duality approach 
described in Sec.~\ref{formalism} and are strongly 
dual~\cite{boyd2004convex}. 

Conversely, the example bounds for periodic films shown in Fig.~4 
are strongly dual only when absorption is near zero. 
Outside this regime, and the associated sharp transition to resonant 
response, all major features, including the half absorption 
plateaus, are seen to be accounted for by the conservation of real 
power, and are thus described by the growth of the radiative 
efficacies (singular values of $\asym{\mathbb{G}^{\text{0}}}$) of 
the two channels with zero in-plane 
wave vector~\cite{molesky2019bounds,kuang2020maximal}.
\\ \\
\textbf{Quasi-Static Regime ($R/\lambda\rightarrow 0,
  ~t/\lambda\rightarrow 0$)} Recalling the conclusions of
Sec.~\ref{toyBounds} B., the simultaneous conservation of real and
reactive power has far-reaching implications for electromagnetic power
transfer when all dimensions of the confining domain are small.  The
analog of the optical theorem for reactive power, \eqref{symOpt}, adds
phase information on top of the maximum polarization magnitude set by
the conservation of real power, \eqref{asymOpt}; and so, when both
constraints are taken into account, \eqref{optProb} captures the fact
it is not always possible to produce a resonant structure given any
single material, of electric susceptibility $\chi$, and a maximal
characteristic size, $R$. In fact, there are rather strict
requirements that must be met for resonant response to be achievable.
Crucially, it must be possible to effectively confine the scattered
electromagnetic field, resulting from the polarization currents
created in the structure by the incident (source) wave, within the
spherical volume.  As validated by the cross section bounds of Fig.~2
and Fig.~3, the only mechanism by which such confinement can be
achieved while also allowing interactions with propagating waves as
$R/\lambda \rightarrow 0$, is the excitation of localized
plasmon-polaritons, occurring for $\re{\chi} = -3$ if the domain is
completely filled with material~\cite{novotny2012principles}.

If $\re{\chi}$ is larger than this value, excluding small deviations
that appear for weak metals between $-2 \gtrsim \re{\chi} \geq -3$,
then, as confirmed by the tiny achievable cross section values,
resonant response with a propagating wave is not possible.  Therefore,
the largest allowed power transfer happens when the material simply
fills the entire domain, and as such, maximal scattering cross
sections exhibit the same susceptibility dependence encountered in
Rayleigh scattering~\cite{tsang2004scattering}. 
Applying \eqref{limResp}, the maximal magnitude of the interaction that can occur in a dielectric structure between the (normalized) 
incident field and the polarization current it excites is
\begin{align}
&\zeta_{\text{Ray}} =
\frac{1}
{\sqrt{\left(\frac{1}{3} + 
\frac{\re{\chi}}{\left|\chi\right|^{2}}\right)^{2} + 
\left(\rho^{_{\mathbb{G}\textbf{N}}}_{1} + 
\frac{\im{\chi}}{\left|\chi\right|^{2}}\right)^{2}}} &\left(\re{\chi}\geq 0\right). 
\label{zetaRayleigh}
\end{align}
with
\begin{align}
\rho^{_{\mathbb{G}\textbf{N}}}_{1} = 
\frac{2}{9}\left(\frac{2\pi R}{\lambda}\right)^{3}
\end{align}
denoting the radiative efficacy of the $\ell =1,~m=\pm1,~0$ 
$\textbf{N}$ polarized channel. 
Employing the power forms given in Sec.~\ref{formalism},  
the scattering cross section of any structure, 
under the above assumptions, must obey the relations 
\begin{align}
\frac{\sigma_{\text{sca}}}{\sigma_{\text{geo}}}
\leq 
\frac{3}{2}\frac{(\rho^{_{\mathbb{G}\textbf{N}}}_{1})^2}
{\left(\frac{1}{3} + 
\frac{\re{\chi}}{\left|\chi\right|^{2}}\right)^{2} + 
\left(\rho^{_{\mathbb{G}\textbf{N}}}_{1} + 
\frac{\im{\chi}}{\left|\chi\right|^{2}}\right)^{2}}
\left(\frac{\lambda}{2\pi R}\right)^{2}
\nonumber \\
\nonumber \\
\frac{\sigma_{\text{abs}}}{\sigma_{\text{geo}}}
\leq 
\frac{3}{2}\frac{\rho^{_{\mathbb{G}\textbf{N}}}_{1}
\left(\im{\chi}/\left|\chi\right|^{2}\right)}
{\left(\frac{1}{3} + 
\frac{\re{\chi}}{\left|\chi\right|^{2}}\right)^{2} + 
\left(\rho^{_{\mathbb{G}\textbf{N}}}_{1} + 
\frac{\im{\chi}}{\left|\chi\right|^{2}}\right)^{2}}
\left(\frac{\lambda}{2\pi R}\right)^{2}
\label{nonResSmallR}.
\end{align}
In stark contrast to $\zeta_{\text{mat}}$, $\zeta_{\text{Ray}}$
decreases for increasing $\re{\chi}$ and has a negligible dependence
on material absorption, $\im{\chi}$. 
Comparing the dashed and full lines of Fig.~2 and Fig.~3, 
particularly Fig.~2 (b) and Fig.~3 (b), the resonance gap between 
these two forms can be quite extreme for realistic dielectrics. 
For example, $\zeta_{\text{mat}}\approx 10^{7}$ for silicon at 
$\lambda=1.5\mu \mathrm{m}$, with $\chi \approx 11 + i~10^{-5}$~
\cite{palik1998handbook}.

As $\re{\chi}$ shifts to increasingly negative values, geometries
supporting localized plasmon--polariton resonances become possible,
and past $\re{\chi} = -3$ cross section limits display resonant
response characteristics. 
The power exchange between the incident field and the generated 
polarization currents is then, asymptotically, restricted to be 
smaller than the ``diluted'' material figure of merit
\begin{align}
&\zeta_{\text{dil}} =
\frac{\left|\chi\right|}{\re{\chi}}
\frac{1}
{\rho^{_{\mathbb{G}\textbf{N}}}_{1} + 
\frac{\im{\chi}}{\left|3\re{\chi}\right|}}   &\left(\re{\chi}\leq -3\right),
\label{zetaMeta}
\end{align}
leading to cross section limits of 
\begin{align}
\frac{\sigma_{\text{sca}}}{\sigma_{\text{geo}}} 
\leq\frac{3}{2}&
\left(\frac{\left|\chi\right|}{\left|\re{\chi}\right|}\right)^{2}
\frac{(\rho^{_{\mathbb{G}\textbf{N}}}_{1})^2}
{
\left(\rho^{_{\mathbb{G}\textbf{N}}}_{1} +
\im{\chi}/\left|3\re{\chi}\right|\right)^{2}
}
\left(\frac{\lambda}{2\pi R}\right)^{2}
\nonumber
\end{align}
\begin{align}
\frac{\sigma_{\text{abs}}}{\sigma_{\text{geo}}}
\leq\frac{3}{2}&
\frac{\left|\chi\right|}{\left|\re{\chi}\right|}
\Bigg(
\frac{\rho^{_{\mathbb{G}\textbf{N}}}_{1}
\left(\im{\chi}/\left|3\re{\chi}\right|\right)}
{
\left(\rho^{_{\mathbb{G}\textbf{N}}}_{1} +
\im{\chi}/\left|3\re{\chi}\right|\right)^{2}
}  
\nonumber \\
&-\frac{(\rho^{_{\mathbb{G}\textbf{N}}}_{1})^2
\left(\left|\chi\right|/\left|\re{\chi}\right|
-1\right)}
{
\left(\rho^{_{\mathbb{G}\textbf{N}}}_{1} +
\im{\chi}/\left|3\re{\chi}\right|\right)^{2}
}
\Bigg)
\left(\frac{\lambda}{2\pi R}\right)^{2}.
\label{resSmallR}
\end{align}
The ``dilution'' modifier is chosen as the form of \eqref{zetaMeta}, 
disregarding the radiative efficacy $
\rho^{_{\mathbb{G}\textbf{N}}}_{1}$, is equivalent to the material 
loss figure of merit $\zeta_{\text{mat}}$ if a ``dilution factor'' 
is introduced to shift $\re{\chi}$ to $-3$. 
That is, considering the low-loss limit $\im{\chi} \ll
\left|\chi\right|$, if it is supposed that the magnitude of
$\re{\chi}$ is rescaled to match the localized resonance condition of
a spherical nanoparticle, then $\zeta_{\text{mat}}\rightarrow
3\left|\chi\right|/\im{\chi}$, which is equal to the expression for 
$\zeta_{\text{dil}}$ in the limit $\rho^{_{\mathbb{G}\textbf{N}}}_{1}
\rightarrow 0$. 
(Due to its connection with the localized plasmon resonance of a 
spherical nanoparticle, the ratio $\left|\chi\right|/\im{\chi}$ is 
commonly encountered in discussing the potential of different 
material options for plasmonic applications~\cite{
maier2007plasmonics,west2010searching}.)

The validity of \eqref{zetaMeta} internally rests on the assumption 
that the wavelength is much larger than any structural feature. 
Since this is also the central criterion for most homogenization 
descriptions of electromagnetic response to be 
applicable~\cite{markel2016maxwell}, it is sensible that material 
structuring limited to tiny domains can, at best, alter  
effective medium parameters~\cite{liu2007description,cai2010optical,
jahani2016all}. 
Equation \eqref{zetaMeta} proves that the implications of this 
picture are universally valid for both scattering and absorption in 
strong metals. 
However, many commonly stated effective medium models also predict 
that there are structures capable of creating effective susceptibility responses more negative than either of the 
constituent materials~\cite{simovski2010electromagnetic,
choy2015effective,mackay2015modern,
petersen2016limitations,lei2017revisiting}. 
For example, the Maxwell--Garnett formula for mono-dispersed 
spherical vacuum inclusions in a background host is
\begin{align}
\chi_{\text{eff}}=\left(1+\chi_{\text{h}}\right)
\frac{1-2f\chi_{h}/\left(3+2\chi_{h}\right)}
{1 + f\chi_{h}/\left(3+2\chi_{h}\right)} -1,
\label{effMedResCM}
\end{align}
where $f$ is the volume filling fraction of the inclusions and
$\chi_{\text{h}}$ is electric susceptibility of the 
host~\cite{merrill1999effective}. 
Based on \eqref{effMedResCM}, using the iterative argument given in 
Ref.~\cite{merrill1999effective}, it should be anticipated that 
low-loss resonant response would be achievable shortly after $\re{
\chi}$ drops below $-1$. 
While cross sections do begin to grow before $\re{\chi} = -3$, it is 
clear that this Maxwell-Garnett condition is not sufficient. 
Dilution via \eqref{effMedResCM} also yields a different dependence 
on material loss-values than that predicted by~\eqref{zetaMeta}, 
with \eqref{effMedResCM} consistently giving slightly larger 
effective losses.

It is also interesting to compare \eqref{nonResSmallR} and
\eqref{resSmallR} with coupled mode descriptions of scattering 
phenomena in the single channel limit~\cite{hamam2007coupled}, 
\begin{align}
\frac{\sigma_{\text{sca}}}{\sigma_{\text{geo}}} &=~ 
6~
\frac{\gamma_{\text{rad}}^{2}}
{\left(\omega - \omega_{\text{res}}\right)^{2}+
\left(\gamma_{\text{rad}} + \gamma_{\text{abs}}\right)^{2}} 
\left(\frac{\lambda}{2\pi R}\right)^{2}
\nonumber \\
\frac{\sigma_{\text{abs}}}{\sigma_{\text{geo}}} &=~ 6~
\frac{\gamma_{\text{rad}}\gamma_{\text{abs}}}
{\left(\omega - \omega_{\text{res}}\right)^{2}+
\left(\gamma_{\text{rad}} + \gamma_{\text{abs}}\right)^{2}}
\left(\frac{\lambda}{2\pi R}\right)^{2},
\label{CMT}
\end{align}
where $\gamma_{\text{rad}}$ and $\gamma_{\text{abs}}$ are the
geometry-specific radiative and absorptive decay rates associated 
with a given resonant mode of frequency $\omega_\text{res}$. 
Up to a missing factor of $4$, which is accounted for by the facts 
that \eqref{nonResSmallR} and \eqref{resSmallR} represent maximum
quantities~\cite{kwon2009optimal,liberal2014least} and that 
scattering cannot occur without 
absorption~\cite{miller2016fundamental}, there is a clear symmetry 
of form between \eqref{nonResSmallR} and \eqref{resSmallR} and 
\eqref{CMT}, provided $\im{\chi} \ll \left|\re{\chi}\right|$ (as 
coupled mode theory requires the assumption of low loss). 
The two sets of expressions agree under the substitutions
\begin{align}
&\gamma_{\text{abs}} \rightarrow 
\im{\chi}/\left|\chi\right|^{2}
\nonumber \\
&\left(\omega -\omega_{\text{res}}\right)^{2}\rightarrow
\left(1/3 + \re{\chi}/\left|\chi\right|^{2}\right)^{2},
\nonumber
\end{align} 
when the system is off resonance, and $$\gamma_{\text{abs}}
\rightarrow \im{\chi}/\left|3\re{\chi}\right|,$$ when the system is 
on resonance; in both situations, $\gamma_{\text{rad}}\rightarrow
\rho^{_{\mathbb{G}\textbf{N}}}_{1}$. 
Since \eqref{nonResSmallR} and \eqref{resSmallR} are bounds, and not 
descriptions of any particular mode, these associations may be 
understood as ``best case'' parameters for what could be achieved in 
any geometry supporting a single mode, and are thus closely linked 
to prior limits based on coupled mode theory~\cite{hamam2007coupled,
verslegers2010temporal,yu2010fundamental,yu2010grating,yu2011angular,
ruan2012temporal}. 
Notably, the comparison precludes any resonant geometry 
from achieving the rate-matching condition of $\gamma_{\text{rad}} =
\gamma_{\text{abs}}$ if it is confined to a small ball. 
Precisely, the only candidate materials are fictitious metals
with $-3 \gtrsim \re{\chi}$ and $\im{\chi}\rightarrow 0$, since the
radiative efficacy $\rho^{_{\mathbb{G}\textbf{N}}}_{1} \rightarrow 0$
with vanishing object size.

This situation, a fictitiously low loss metallic nanoparticle, is 
also the most relevant condition under which the bounds
asymptotically reach arbitrarily large values. 
However, as we have discussed in Ref.~\cite{molesky2019bounds} in 
the context of angle-integrated planewave absorption, unbounded 
growth requires saturation of an unbounded number of angular 
momentum $\ell$ indices (radiation channels). 
For any particular $\ell$, saturation is approximately achieved as 
$R\rightarrow 0$ when $\rho_{\ell}^{_{\mathbb{G}\textbf{N}}},~
\rho_{\ell}^{_{\mathbb{G}\textbf{M}}} \gtrsim \im{\chi}/\left|3\re{
\chi}\right|$. 
Therefore, the relation between the radiative efficacies and the 
angular momentum number $\ell$, with 
\begin{align}
\rho^{_{\mathbb{G}\textbf{N}}}_{\ell}\left(R\right) =& \frac{\pi
\left(kR\right)^{2}}{4} 
\nonumber \\
&\Bigg[
\frac{\ell+1}{2\ell+1}\left(J_{\ell-\frac{1}{2}}^{2}\left(kR\right)
-
J_{\ell+\frac{1}{2}}\left(kR\right)J_{\ell-\frac{3}{2}}\left(kR\right)\right)
\nonumber \\ &\hspace{-0.2in}
+\frac{\ell}{2\ell+1}\left(J_{\ell+\frac{3}{2}}^{2}\left(kR\right) -
J_{\ell+\frac{1}{2}}\left(kR\right)J_{\ell+\frac{5}{2}}\left(kR\right)\right)\Bigg],
\nonumber \\ \rho^{_{\mathbb{G}\textbf{M}}}_{\ell}\left(R\right) =&
\frac{\pi\left(kR\right)^{2}}{4}\left(J_{\ell+\frac{1}{2}}^{2}\left(kR\right)
- J_{\ell-\frac{1}{2}}\left(kR\right)
J_{\ell+\frac{3}{2}}\left(kR\right)\right),
\label{singVals}
\end{align}
imparted through the cylindrical Bessel functions, with
\begin{align}
J_{\ell+p}\left(kR\right)<
\frac{1}{\Gamma\left(\ell+p +1\right)}\left(\frac{kR}{2}\right)^{\ell +p}
\label{besselEll}
\end{align}
for $kR<\sqrt{8\Gamma\left(\ell + p +3\right)}$, implies that so long
as real power is conserved, bounds on cross section enhancement
exhibit sublogarithmic growth with vanishing material loss, 
$\im{\chi}\rightarrow 0$. 
(Proof of this statement, in all important regards, follows from the 
derivation given in Ref.~\cite{molesky2019bounds}. 
The stated inequality follows from the power series of the 
cylindrical Bessel functions~\cite{NIST:DLMF}.)  
As seen in the supporting figures, Fig.~2 and Fig.~3, in practice 
this scaling behavior is of little consequence.

For periodic films (e.g., gratings, photonic crystals, and
metasurfaces), the central feature of \eqref{optProb} absent from the
models of Ref.~\cite{molesky2019bounds} and
Ref.~\cite{kuang2020maximal} is the initial suppression and sharp 
onset of resonant absorption for thin dielectrics. 
The above findings for compact domains, and practical experience, 
both suggest that the existence of such a critical 
thickness (depending of the magnitude of $\re{\chi}$) for dielectric 
materials is reasonable. 
However, the dependence of this thickness threshold on the
period of the system, physically associated with the presence of
leaky-mode resonances~\cite{joannopoulosphotonic}, 
is perhaps less expected. 
The origin of the relation traces to the properties of 
$\mathbb{G}^{\text{0}}$ for an extended (infinite) system.  
Crossing over the light line boundary between propagating and 
evanescent waves, there are vectors within the basis described 
in Sec.~\ref{appendix}, $\left|\textbf{X}\right> =
\left|\hat{\textbf{RX}}^{\left(-\right)}\left(\text{k}_{p}\right)
\right>$, that allow $\left<\textbf{X}\right|
\sym{\mathbb{G}^{\text{0}}}\left|\textbf{X}\right>$ to be arbitrarily
negative and $\left<\textbf{X}\right|
\asym{\mathbb{G}^{\text{0}}}\left|\textbf{X}\right> = 0$. 
These characteristics allow reactive power conservation, 
\eqref{symOpt}, to be trivially satisfied for any possible 
polarization current. 
However, when a finite period is imposed on the system, modes 
arbitrarily near the light line are not allowed, and in turn, the 
necessity of conserving reactive power may imply that resonant 
response is not possible for particular values of $t$ and $\chi$. 
As exemplified in Fig.~4, knowledge of the critical thickness at 
which such leaky modes can be supported for a given period and 
material may be of substantial benefit to the design of large scale
metasurfaces~\cite{wu2019tunable, lin2019topology,jin2020inverse}.
\\ \\
\noindent
\textbf{Wavelength Scale Regime ($R/\lambda\gtrsim 0.1,
~t/\lambda\gtrsim 0.01$)} 
For boundary radii approaching wavelength size, the applicability of 
the quasi-static results for spherical confining region quoted under 
the previous subheading becomes increasingly tenuous. 
The growth of planewave amplitude coefficients into angular momentum 
numbers (channels) beyond $\ell = 1$ opens the possibility of
utilizing a wider range of wave physics (e.g., leaky and guided
resonances~\cite{li2019leaky,lee2019band}), and correspondingly, 
reactive power conservation (resonance creation) becomes a weaker 
requirement. 

These factors lead to a more intricate interplay between the two 
power constraints, causing the sharp jumps observed for dielectrics 
in Fig.~2 and Fig.~3, which manifest, mechanically, in rapid changes 
to the  properties of the scattering $\mathbb{T}$ operator 
constraint  relations, especially \eqref{symOpt} applied to 
dielectric materials. 
The behavior is first observed in the $\ell=1$ channel, with the 
initial peaks in Fig.~2(b) and Fig.~3(b) following closely after the 
half wavelength condition
$$\left(\text{min}~r\right) ~\ni~ 
\partial j_{1}\left(\sqrt{\re{\chi}}~\frac{2\pi r}{\lambda}\right)/\partial r = 0,$$ 
and the second peaks occurring near the full wavelength condition,
$$\left(\text{min}~r\right) ~\ni~ j_{1}\left(\sqrt{\re{\chi}}~
\frac{2\pi r}{\lambda}\right) = 0.$$
This second criterion is also the approximate resonance location 
for a homogeneous dielectric sphere of index $\sqrt{\chi}$~
\cite{newton2013scattering}, the spherical analogs of the 
Fabry--Perot condition~\cite{yariv2006photonics}, making its 
appearance consistent with the Rayleigh response predictions of 
\eqref{zetaRayleigh}. 
At the same time, as previously remarked, the inflation of the 
boundary also increases the radiative efficacy of each channel as 
described by \eqref{singVals} (further discussed in Ref.
~\cite{molesky2019bounds} and Sec.~\ref{priorWork}).
Via the connection of $\asym{\mathbb{U}_{\ell}}$ to $
\asym{\mathbb{G}^{\text{0}}_{\ell}}$, \eqref{uDef}, this causes the 
conservation of real power to become a more restrictive constraint 
for generating strong polarization currents throughout the volume 
available to structuring. 
(That is, maintaining the ratio of $\im{\left<\textbf{S}^{(1)}|
\textbf{T}\right>}$ and $\left<\textbf{T}\right|\asym{\mathbb{U}}
\left|\textbf{T}\right>$ becomes increasingly restrictive, while 
maintaining the ratio of $\re{\left<\textbf{S}^{(1)}|\textbf{T}
\right>}$ and $\left<\textbf{T}\right|\sym{\mathbb{U}}\left|
\textbf{T}\right>$ becomes increasingly simple.) 
Subsequently, rather than completely releasing to an enhancement 
value approaching $\zeta_{\text{mat}}$, the bound slips and catches. 

For periodic films, all features seen at both wavelength and 
ray optic thickness scales are fully accounted for by the 
conservation of real power and the associated radiative efficacies of
$\asym{\mathbb{G}^{\text{0}}}$ at normal incidence.  Detailed
discussions of these quantities are given in
Ref.~\cite{molesky2019bounds} and Ref.~\cite{kuang2020maximal}. 
The most interesting result, that the bounds plateau for film 
thickness between roughly $t/\lambda = 10^{-4}$ and 
$t/\lambda = 10^{-2}$, is caused by the fact that at these 
thicknesses only a single symmetric mode is bright. 
Accordingly, the power radiated to the far-field by the
excited polarization current can not be canceled, and only half 
of the total power of the incident wave can be extracted through
absorption~\cite{molesky2019bounds}.  
\\ \\ 
\textbf{Ray Optics Regime ($R/\lambda \gg 1, ~t/\lambda 
\gg 10^{-1}$)} 
In the large boundary limit, achievable scattering interactions in 
any given channel are increasingly dominated by the conservation of 
real power through the growth of radiative losses. 
Correspondingly, the dash bounds, calculated by
asserting only that the sum of the scattered and absorbed power must
not exceed the power drawn from the incident beam, coincide with 
those arising from total power conservation to increasingly good 
accuracy.
Making this reduction, limits for either cross section enhancement
quantity become largely congruous to the angle-integrated absorption
bounds given in Ref.~\cite{molesky2019bounds}. 
The planewave expansion coefficients of \eqref{planeWave} exhibit 
exactly the same per-channel characteristics considered in that 
article, and so, the same asymptotic behavior is encountered. 
Regardless of the selected susceptibility $\chi$, for a sufficiently 
large radius, each of the power objectives described in 
Sec.~\ref{formalism} begins to scale as the geometric cross section 
of the bounding sphere. 
For absorption, this leads to a value equal to the geometric cross 
section of the confining ball, $\pi R^{2}$. 
For extinction and scattering, a value of $4\pi R^{2}$ is found, two 
times larger than what would be expected based on the extinction 
paradox~\cite{zakowicz2002extinction,berg2011new}. 
The genesis of this additional factor is presently unknown, and
investigation of the properties of the optimal polarization current 
of these curious results merits further study.

\section{Summary Remarks}\label{summary}

The ability of metals and polaritonic materials to confine light in
subwavelength volumes without the need for any other surrounding
structure (plasmon--polaritons~\cite{williams2008highly,
neutens2009electrical}), coupled with the variety of geometric wave
effects achievable in dielectric media (band
gaps~\cite{foresi1997photonic,chen2016multi}, index
guiding~\cite{li2017controlling,le20163d}, topological
states~\cite{ozawa2019topological,khanikaev2017two}), rest as the
bedrock of contemporary photonic design. 
Yet, the relative abilities of these two overarching approaches for 
controlling light--matter interactions remains a widely studied 
topic~\cite{noginov2018miniature,khurgin2018relative,
ballarini2019polaritonics}. 
The broad strokes are well established.
The possibility of subwavelength confinement and large field 
enhancements offered by metals is offset by the fact these effects 
are fundamentally linked to substantial material
loss~\cite{khurgin2018relative}. 
Through interference, dielectric architectures may also confine and 
intensify electromagnetic fields, and can do so without large 
accompanying material absorption~\cite{lin2016cavity}; but, 
accessing this potential invariably requires larger domains and more 
complex structures. 
While comparisons within rigidly defined subclasses have been
made~\cite{liu2016fundamental}, the merit of a particular method for 
a particular design challenge is almost always an open question. 
As with the rising need for limits in computational approaches 
highlighted in the introduction, a central driver of debate is the 
lack of concrete (pertinent) knowledge of what is possible, beyond 
qualitative arguments.

We believe that the simple instructive cross section examples 
shown in Sec.~\ref{applications} are compelling evidence that 
the generation of bounds based on constraints derived 
from the $\mathbb{T}$ operator and Lagrange
duality offers a path towards progress; and that  
by translating this method beyond the spectral basis employed
here, onto a completely geometry agnostic numerical algorithm, 
it will be possible to analyze the relative 
trade offs associated with various kinds of optical devices. 
Through bound calculations varying material and domain parameters, 
the significance of different design elements from the perspective 
of device performance should be ascertainable in a number of 
technologically relevant areas. 
The basic scattering interaction quantities given in 
Sec.~\ref{formalism} lie at the core of engineering the radiative 
efficacy of quantum emitters~\cite{lu2017dynamically,
davoyan2018quantum,crook2020purcell}, resonant response of 
cavities~\cite{lin2016enhanced,liu2017quantum,wang2018maximizing},
design characteristics of
metasurfaces~\cite{kruk2017functional,groever2017meta,lewi2019thermally}, and the efficacy of light 
trapping~\cite{yu2010grating,sheng2012light}
devices and luminescent~\cite{zalogina2018purcell,valenta2019nearly}
and fluorescent~\cite{li2017plasmon,simovski2019point} sources. 
They are also central building blocks of quantum and nonlinear 
phenomena like F\"{o}rster energy 
transfer~\cite{cortes2018fundamental}, Raman 
scattering~\cite{michon2019limits}, and frequency
conversion~\cite{lin2016cavity}.

As seen in Sec.~\ref{applications}, relations \eqref{symOpt} and 
\eqref{asymOpt} are amenable to numerical evaluation under realistic 
photonic settings (for practical domain sizes and materials) and 
sufficiently broad to provide both quantitative guidance and 
physical insights: as the size of an object interacting with a 
planewave grows, there is a transition from the volumetric (or super 
volumetric) scaling characteristic of subwavelength objects to the 
geometric cross section dependence characteristic of ray optics; 
critical sizes exist below which it is impossible to create 
dielectric resonances; material loss dictates achievable 
interactions strengths only once it becomes feasible to achieve 
resonant response and significant coupling to the incident field. 

Several generalizations of the formalism should be possible. 
First, there is an apparent synergy with the work of Angeris, 
Vu{\v{c}}kovi{\'c} and Boyd~\cite{angeris2019computational} for 
inverse design applications. 
The optimal vectors found using \eqref{optProb} provide intuitive 
target fields. 
Second, following the arguments given in the work of 
Shim el al.~\cite{shim2019fundamental} it would seem that 
\eqref{optProb} can be further enlarged to include finite bandwidth 
dispersion information, accounting for the full analytic features of 
the electric susceptibility $\chi\left(\omega\right)$. 
Finally, by combining the respective strengths of both classes of 
materials, hybrid metal-dielectric structures offer the potential of 
realizing more performant devices. 
The generalization of \eqref{optProb} to incorporate multiple 
material regions (multi-region 
scattering~\cite{molesky2020fundamental}) as an aid to these 
efforts stands as an important direction of ongoing study. 
As we have stated earlier, as the method relies only on scattering 
theory, almost all lines of reasoning we have presented apply 
equally to acoustics, quantum mechanics, and other wave physics. 

\section{Acknowledgments}
\noindent
This work was supported by the National Science Foundation under
Grants No. DMR-1454836, DMR 1420541, DGE 1148900, the Cornell Center
for Materials Research MRSEC (award no. DMR1719875), and the Defense
Advanced Research Projects Agency (DARPA) under Agreement
No. HR00112090011. 
The views, opinions and/or findings expressed herein are those of 
the authors and should not be interpreted as representing the official views or policies of any institution. 
We thank Prashanth S. Venkataram, Jason Necaise, and Prof. Shanhui
Fan for useful comments.

\section{Appendix}\label{appendix}

\subsection{Numerical Stability of the Arnoldi Processes}

With perfect numerical accuracy, the convergence of 
$\mathbb{A}_{\ell}$ is guaranteed in a finite number of 
iterations. The strictly diagonal elements of each 
$\mathbb{U}_{\ell}$ matrix, $\mathbb{V}^{\dagger -1}_{\ell}$
, remain constant while the off diagonal coupling coefficients 
introduced by $\sym{\mathbb{G}^{\text{0}\dagger}_{\ell}}$ 
gradually decay with every iterations. Thus, at a certain point, the 
diagonal $\mathbb{V}^{\dagger -1}_{\ell}$ entries eventually 
overwhelm all other contributions, terminating the 
$\mathbb{U}_{\ell}$ matrix. (As the magnitude of the 
susceptibility considered increases, $\mathbb{V}^{\dagger -1}$ 
shrinks and more Arnoldi iterations are required.)

Still, there are pitfalls that must be avoided when numerically
implementing an Arnoldi iteration, caused by the singularity of the 
outgoing $\textbf{N}$ waves at the origin. The issue is illustrated 
by considering the image of $\textbf{RN}$ under 
$\mathbb{G}^{\text{0}}$ (\ref{GreenExpand}), with
\begin{equation}
\textbf{N}_{\ell,m}\left(\textbf{x}\right) 
= \frac{\sqrt{\ell\left(\ell+1\right)}}{r} 
\text{h}_{\ell}^{\left(1\right)}\left(r\right)
\text{A}_{\ell ,m}^{\left(3\right)} + 
\frac{\partial\left(x~\text{h}_{\ell}^{\left(1\right)}
\left(\text{r}\right)\right)}{r} 
\text{A}_{\ell,m}^{\left(2\right)},
\label{outN}
\end{equation}
using the normalized vector spherical harmonics as described in 
Ref~\cite{KristenssonEMwaves}.
Near the origin, $r\rightarrow0$, the leading order radial dependencies of \eqref{regSols} and \eqref{outN} are 
\begin{align}
\textbf{RN}_{\ell,m} =& \left(\frac{\ell+1}{(2\ell+1)!!}\left(r\right)^{\ell-1} + 
\Ord{ r^{\ell+1}}\right)\text{A}_{\ell ,m}^{\left(2\right)} + \nonumber\\
&\left(\frac{\sqrt{\ell(\ell+1)}}{(2\ell+1)!!}r^{\ell-1} + 
\Ord{r^{\ell+1}}\right)\text{A}_{\ell ,m}^{\left(3\right)}, 
\label{asympRgN}
\\
\textbf{N}_{\ell,m} =& \left(\frac{i \ell(2\ell-1)!!}{r^{\ell+2}} + \Ord{r^{-\ell}}\right)\text{A}^{\left(2\right)}_{\ell ,m} + \nonumber\\ 
&\left(\frac{-i(2\ell-1)!!\sqrt{\ell(\ell+1)}}{r^{\ell+2}} + \Ord{r^{-\ell}}\right) 
\text{A}_{\ell ,m}^{\left(3\right)}.
\label{asympN}
\end{align}
From \eqref{GreenExpand}, the image of $\textbf{RN}_{\ell,m}$ under 
the Green function restricted to a spherical 
domain with radius $R$ is
\begin{equation}
\mathbb{G}^{\text{0}}\textbf{RN} = 
\textbf{RN}(\textbf{r}) \textbf{RN}_{co}(\textbf{r}) + 
\textbf{N}(\textbf{r}) \textbf{N}_{co}(\textbf{r}) - 
\textbf{RN}(\textbf{r})~\textbf{A}^{\left(3\right)}_{\ell ,m}
\end{equation}
where the final term is the $\delta$-function contribution, and the 
$\textbf{RN}_{co}(\textbf{r})$ and $\textbf{N}_{co}(\textbf{r})$ 
terms are given by 
\begin{align}
\textbf{N}_{co}(\textbf{r}) &= 
i \iint_{\Omega'} \int_0^{r} r'^2 
\textbf{RN}(\textbf{r'})\textbf{RN}(\textbf{r'}) 
\text{d}r' \text{d}\Omega', 
\nonumber \\
\textbf{RN}_{co}(\textbf{r}) &= 
i \iint_{\Omega'} \int_r^R r'^2 
\textbf{N}(\textbf{r'}) 
\textbf{RN}(\textbf{r'}) 
\text{d}r' \text{d}\Omega'.
\end{align}
Exploiting the orthogonality of the vector spherical harmonics, 
the leading radial order for 
$\textbf{N}_{co}(\textbf{r})$ is $r^{2\ell+1}$. Therefore, the
$\textbf{N}(\textbf{r}) \textbf{N}_{co}(\textbf{r})$ term has a 
leading radial order of $r^{\ell-1}$, the same as the starting vector 
$\textbf{RN}(\textbf{r})$.
At first sight, $\textbf{RN}_{co}(\textbf{r})$ is more troubling. 
The dominate radial orders are $r'^{-\ell}$ for 
$r'^2\textbf{N}(\textbf{r'})$ and $r'^{\ell-1}$ for 
$\textbf{RN}(\textbf{r'})$. Thus, it would seem that the integrand 
has an $r'^{-1}$ dependence, which would result in a 
logarithmic divergence at the origin. A more careful 
consideration, however, shows that the leading order terms 
from $\textbf{A}^{\left(2\right)}_{\ell ,m}$ and 
$\textbf{A}^{\left(3\right)}_{\ell ,m}$ cancel as
\begin{align}
\textbf{RN}_{co}(\textbf{r}) =& 
~i \iint_{\Omega'}\bigg( \frac{i\ell(\ell+1)}{2\ell+1} r'^{-1} 
\text{A}^{\left(2\right)}_{\ell ,m}
\cdot\text{A}^{\left(2\right)}_{\ell ,m} - 
\nonumber\\
&\frac{i\ell(\ell+1)}{2\ell+1} r'^{-1} 
\text{A}^{\left(3\right)}_{\ell ,m}\cdot
\text{A}^{\left(3\right)}_{\ell ,m} + 
\Ord{r'} \text{d}r' \bigg)\text{d}\Omega 
\nonumber \\
=&~\Ord{r^2}.
\end{align}
The key to this cancellation is the ratio of the 
$\text{A}_{\ell ,m}^{\left(2\right)}$ and
$\text{A}^{\left(3\right)}_{\ell ,m}$ terms, $\sqrt{(\ell+1)/\ell}$.
So long as this ratio is maintained, the $\textbf{RN}_{co}$ factor 
does not generate logarithmic contributions, and in turn this causes 
the leading order ratio to remain intact under the further action of 
$\mathbb{G}^{\text{0}}$. By insuring that this does in fact occur, 
the Arnoldi process may continue to stably iterate until convergence 
is achieved. Consider any vector 
\begin{equation}
\textbf{P} = p\left(r^{\ell-1} 
\text{A}_{\ell ,m}^{\left(2\right)} + 
\sqrt{\frac{\ell}{\ell+1}}r^{\ell-1} 
\text{A}_{\ell ,m}^{\left(3\right)}\right),
\end{equation}
where $p$ is a constant. 
($\textbf{RN}_{\ell,m}$ are vectors of this form.) 
The image under of this vector under $\mathbb{G}^{\text{0}}$ is 
\begin{align}
\mathbb{G}^{\text{0}} \textbf{P} =& ~\textbf{RN}_{\ell,m}
\left(\textbf{r}\right)
\textbf{RN}^{_\textbf{P}}_{co}\left(\textbf{r}\right) + 
\textbf{N}_{\ell,m}\left(\textbf{r}\right) 
\textbf{N}^{_\textbf{P}}_{co}(\textbf{r}) - 
\nonumber \\
&p~\sqrt{\frac{\ell}{\ell+1}} r^{\ell-1} 
\text{A}^{\left(3\right)}_{\ell ,m}, 
\label{eqnGP}
\end{align}
with
\begin{align}
\textbf{RN}^{_\textbf{P}}_{co}(\textbf{r}) =& 
~p\int_{r}^{R} \Big(
\frac{i \ell \left(2\ell-1\right)!!}{r'^\ell} r'^{\ell-1} -
\nonumber \\
&\frac{i\ell\left(2\ell-1\right)!!}{r'^\ell} r'^{\ell-1}  
+ \mathcal{O}\left(r'\right) \Big) \text{d}r' 
\nonumber\\
=&~ C\left(R\right) + \mathcal{O}\left(r^2\right),
\end{align}
$C\left(R\right)$ a constant of $r$ coming from the fixed upper integration limit $R$, and
\begin{align}
\textbf{N}^{_\textbf{P}}_{co}(\textbf{r}) &= 
ip\int_0^{r}\left( \frac{2\ell+1}{(2\ell+1)!!}r'^{2\ell}  
+ \Ord{r'^{2\ell+3}} \right) 
\text{d}r 
\nonumber \\
&= \frac{i p}{(2\ell+1)!!} r^{2\ell+1} + \Ord{r^{\ell+3}}.
\end{align}
Substituting back into (\ref{eqnGP}) then gives
\begin{align}
\mathbb{G}^{\text{0}} \textbf{P} =& 
\left(-\frac{\ell p}{2\ell+1} r^{\ell-1} + 
\Ord{r^{\ell+1}}\right)\text{A}_{\ell ,m}^{\left(2\right)} +
\nonumber \\
&\left(-\frac{\ell p}{2\ell+1}\sqrt{\frac{\ell}{\ell+1}} 
r^{\ell-1} + \Ord{r^{\ell+1}} \right) 
\text{A}_{\ell ,m}^{\left(3\right)} +
\nonumber \\ 
&C\left(R\right) \textbf{RN}(\textbf{r}).
\end{align}
Hence, as anticipated, all components retain a 
$\sqrt{\ell/(\ell+1)}$ ratio. 
By induction, this argument extends to every step of the Arnoldi 
process, generating vectors well behaved at the origin.

In implementation, care must be taken not to let numerical error 
push this component ratio away from $\sqrt{\ell/(\ell+1)}$ at any 
step. (Otherwise, the logarithmic divergence will quickly 
destabilize new image vectors.) This precludes the use of spatial 
discretization based representations, since for finite grids 
discretization error is inevitable and leads to a rapidly growing 
instability. We have circumvented this issue by 
representing the radial dependence of the Green function and 
Arnoldi vectors by polynomials (Taylor series). 
For larger domain sizes, this approach demands a high level of 
numeric precision, and so, the Python arbitrary precision 
floating-point arithmetic package \emph{mpmath} was used in all 
calculations~\cite{mpmath}. When determining the image of a vector 
under the Green function, the tiny coefficient of $r'^{-1}$ due to 
numerical errors from the finite Taylor series and set 
floating-point precision were explicitly truncated (ignored). 
With sufficiently high precision and representation order 
the Arnoldi process can be performed stably and accurately up to 
convergence of each $\mathbb{U}_{\ell}$ matrix. 
Much of the difficulty, and inefficiency, associated with this 
method stems from working in spherical coordinate, 
which are inherently ill defined at the origin. 

\subsection{Radiative (Fourier) Basis for Films}

Working in Cartesian coordinates, consider a planar slab of thickness
$t$ with normal direction $\uv{z}$. Let $\vec{k}_{p}$ denote the
in-plane wave vector with corresponding spatial vector
$\vec{x}_{p}$. Following our notation of including an extra factor of
$k^{2}$ as mentioned above, the Fourier representation of the Green
function given in Ref.~\cite{ tsang2004scattering} below the light
line ($k_{p} / k < 1$) is given by
\begin{align}
&\mathbb{G}^{\text{0}}\left(\vec{x},\vec{y}\right)= -\int\limits_{\text{Y}}\delta\left(\vec{x}-\vec{y}\right)\mathbb{P}_{\uv{z}} + i\int\limits_{0}^{1} d\vec{k}_{p} \int\limits_{\text{Y}}
\nonumber \\
&
\begin{cases}
\vec{O}^{\left(+\right)}_{\vec{kp}}\left(\vec{x}\right)\otimes\vec{O}^{\left(+\right)}_{\vec{kp}}\left(\vec{y}\right) + 
\vec{X}^{\left(+\right)}_{\vec{kp}}\left(\vec{x}\right)\otimes\vec{X}^{\left(+\right)}_{\vec{kp}}\left(\vec{y}\right) & x_{z} > y_{z}
\\
\vec{O}^{\left(-\right)}_{\vec{kp}}\left(\vec{x}\right)\otimes\vec{O}^{\left(-\right)}_{\vec{kp}}\left(\vec{y}\right) + 
\vec{X}^{\left(-\right)}_{\vec{kp}}\left(\vec{x}\right)\otimes\vec{X}^{\left(-\right)}_{\vec{kp}}\left(\vec{y}\right)&x_{z} < y_{z}
\end{cases}
\label{greenPlanarSubLight}
\end{align}
with 
\begin{align}
\vec{O}^{\left(\pm\right)}_{\vec{kp}}\left(\vec{x}\right) &= 
\frac{e^{i\vec{k}^{\left(\pm\right)}\vec{x}}}{2\pi}
\frac{\uv{o}}{\sqrt{2k_{z}}},
\nonumber \\
\vec{X}^{\left(\pm\right)}_{\vec{kp}}\left(\vec{x}\right) &=
\frac{e^{i\vec{k}^{\left(\pm\right)}\vec{x}}}{2\pi}
\frac{\mp k_{z}\uv{k}_{p}+k_{p}\uv{z}}{\sqrt{2k_{z}}}
,
\label{greenWavesSubLight}
\end{align}
and $\vec{o}=\left(k_{y}\uv{x}-k_{x}\uv{y}\right)$.
The four terms are, respectively, the upward traveling and downward 
traveling ordinary and extraordinary waves~\cite{chen1983theory}. 
In \eqref{greenPlanarSubLight} and \eqref{greenWavesSubLight} all 
wave vectors are normalized by the standard wave vector $k = 2\pi
/\lambda$, and all spatial coordinates are multiplied by $k$; 
$k_{z}$ is defined to be the positive square root 
$k_{z} = \sqrt{1-k_{p}^{2}}$; $k^{\left(\pm\right)} = \vec{k}_{p} 
\pm k_{z} \uv{z}$; and when a wave appears on the right of a 
$\otimes$ symbol complex conjugation is implied. 
Anti-symmetrizing, the skew-Hermitian component is
\begin{align}
\asym{\mathbb{G}^{\text{0}}}\left(\vec{x},\vec{y}\right) =& 
\int\limits_{k_{p}\leq 1}\int_{\text{Y}}d\vec{k}_{p} \sum_{\left(\pm\right)}
\nonumber \\
&\rho^{_{\mathbb{G}\textbf{O}^{\left(\pm\right)}}}_{k_{p}}
\left(\uv{RO}^{\left(\pm\right)}_{\vec{k}_{p}}\left(\vec{x}\right)\otimes 
\uv{RO}^{\left(\pm\right)}_{\vec{k}_{p}}\left(\vec{y}\right)\right) +
\nonumber \\
&\rho^{_{\mathbb{G}\textbf{X}^{\left(\pm\right)}}}_{k_{p}}
\left(\uv{RX}^{\left(\pm\right)}_{\vec{k}_{p}}\left(\vec{x}\right)\otimes 
\uv{RX}^{\left(\pm\right)}_{\vec{k}_{p}}\left(\vec{y}\right)\right),
\label{asymGreenPlanar}
\end{align}
with 
\begin{align}
\uv{RO}^{\left(+\right)}_{\vec{k}_{p}}\left(\vec{x}\right) &= 
\frac{e^{i\vec{k}_{p}\vec{p}}}{2\pi}
~\frac{\sqrt{2}~\text{cos}\left(k_{z}z\right)\uv{o}}
{\sqrt{t\left(1+\text{sinc}\left(k_{z} t\right)\right)}}
\nonumber \\
\uv{RO}^{\left(-\right)}_{\vec{k}_{p}}\left(\vec{x}\right) &= 
\frac{e^{i\vec{k}_{p}\vec{p}}}{2\pi}
\frac{\sqrt{2}~\text{sin}\left(k_{z}z\right)\uv{o}}
{\sqrt{t\left(1-\text{sinc}\left(k_{z} t\right)\right)}},
\nonumber \\
\uv{RX}^{\left(+\right)}_{\vec{k}_{p}}\left(\vec{x}\right) &= 
\frac{e^{i\vec{k}_{p}\vec{p}}}{2\pi}
\frac{\sqrt{2}\left(k_{z}~\text{cos}\left(k_{z}z\right)\uv{k}_{p} 
- i k_{p}~\text{sin}\left(k_{z}z\right)\uv{z}\right)}
{\sqrt{t\left(1+\left(k_{z}^{2}-k_{p}^{2}\right)
\text{sinc}\left(k_{z}t\right)\right)}},
\nonumber \\
\uv{RX}^{\left(-\right)}_{\vec{k}_{p}}\left(\vec{x}\right) &= 
\frac{e^{i\vec{k}_{p}\vec{p}}}{2\pi}
\frac{\sqrt{2}\left(k_{z}~\text{sin}\left(k_{z}z\right)\uv{k}_{p} 
+i k_{p}~\text{cos}\left(k_{z}z\right)\uv{z}\right)}
{\sqrt{t\left(1-\left(k_{z}^{2}-k_{p}^{2}\right)
\text{sinc}\left(k_{z}t\right)\right)}},
\label{radPlanarSubLight}
\end{align}
and
\begin{align}
\rho^{_{\mathbb{G}\textbf{O}^{\left(\pm\right)}}}_{k_{p}} &= 
\frac{t}{4k_{z}}\left(1\pm\text{sinc}\left(k_{z}t\right)\right),
\nonumber \\
\rho^{_{\mathbb{G}\textbf{X}^{\left(\pm\right)}}}_{k_{p}} &= 
\frac{t}{4k_{z}}\left(1\pm\left(k_{z}^{2}-k_{p}^{2}\right)\text{sinc}\left(k_{z}t\right)\right).
\label{planarRadEff}
\end{align}
Above the light line ($k_{p} / k > 1$), the Green function becomes
Hermitian and \eqref{greenWavesSubLight} is replaced by
\begin{align}
&\mathbb{G}^{\text{0}}\left(\vec{x},\vec{y}\right)= -\int\limits_{\text{Y}}\delta\left(\vec{x}-\vec{y}\right)\mathbb{P}_{\uv{z}} + \int\limits_{1}^{\infty} d\vec{k}_{p} \int\limits_{\text{Y}}
\nonumber \\
&
\begin{cases}
\vec{O}^{\left(+\right)}_{\vec{ke}}\left(\vec{x}\right)\otimes\vec{O}^{\left(-\right)}_{\vec{ke}}\left(\vec{y}\right) + 
\vec{X}^{\left(+\right)}_{\vec{ke}}\left(\vec{x}\right)\otimes\vec{X}^{\left(-\right)}_{\vec{ke}}\left(\vec{y}\right) & x_{z} > y_{z}
\\
\vec{O}^{\left(-\right)}_{\vec{ke}}\left(\vec{x}\right)\otimes\vec{O}^{\left(+\right)}_{\vec{ke}}\left(\vec{y}\right) + 
\vec{X}^{\left(-\right)}_{\vec{ke}}\left(\vec{x}\right)\otimes\vec{X}^{\left(+\right)}_{\vec{ke}}\left(\vec{y}\right)&x_{z} < y_{z}
\end{cases}
\label{greenPlanarSubLightSupLight}
\end{align}
where 
\begin{align}
\vec{O}^{\left(\pm\right)}_{\vec{ke}}\left(\vec{x}\right) &= 
\frac{e^{i\vec{k}_{p}\vec{x}_{p} \mp k_{z}x_{z}}}{2\pi}
\frac{\uv{o}}{\sqrt{2k_{z}}},
\nonumber \\
\vec{X}^{\left(\pm\right)}_{\vec{ke}}\left(\vec{x}\right) &=
\frac{e^{i\vec{k}_{p}\vec{x}_{p} \mp k_{z}x_{z}}}{2\pi}
\frac{\mp ik_{z}\uv{k}_{p}+k_{p}\uv{z}}{\sqrt{2k_{z}}}
,
\label{greenWavesSupLight}
\end{align}
and $k_{z}$ has been redefined as $k_{z} = \sqrt{k_{p}^{2}-1}$. 
Although $\asym{\mathbb{G}} = 0$, by analytic extension a ``radiative'' basis (symmetric and anti-symmetric combinations of the two ordinary and extraordinary waves) is given by 
\begin{align}
\uv{RO}^{\left(+\right)}_{\vec{k}_{p}}\left(\vec{x}\right) &= 
\frac{e^{i\vec{k}_{p}\vec{p}}}{2\pi}
~\frac{\sqrt{2}~\text{cosh}\left(k_{z}z\right)\uv{o}}
{\sqrt{t\left(1+\text{sinch}\left(k_{z} t\right)\right)}}
\nonumber \\
\uv{RO}^{\left(-\right)}_{\vec{k}_{p}}\left(\vec{x}\right) &= 
\frac{e^{i\vec{k}_{p}\vec{p}}}{2\pi}
\frac{\sqrt{2}i~\text{sinh}\left(k_{z}z\right)\uv{o}}
{\sqrt{t\left(\text{sinch}\left(k_{z} t\right)-1\right)}},
\nonumber \\
\uv{RX}^{\left(+\right)}_{\vec{k}_{p}}\left(\vec{x}\right) &= 
\frac{e^{i\vec{k}_{p}\vec{p}}}{2\pi}
\nonumber \\
&\frac{\sqrt{2}\left(k_{z}~\text{cosh}\left(k_{z}z\right)\uv{k}_{p} 
-i k_{p}~\text{sinh}\left(k_{z}z\right)\uv{z}\right)}
{\sqrt{t\left(\left(k_{p}^{2} + k_{z}^{2}\right)
\text{sinch}\left(k_{z}t\right) -1\right)}},
\nonumber \\
\uv{RX}^{\left(-\right)}_{\vec{k}_{p}}\left(\vec{x}\right) &= 
\frac{e^{i\vec{k}_{p}\vec{p}}}{2\pi}
\nonumber \\
&\frac{\sqrt{2}\left(k_{z}~\text{sinh}\left(k_{z}z\right)\uv{k}_{p} 
-i k_{p}~\text{cosh}\left(k_{z}z\right)\uv{z}\right)}
{\sqrt{t\left(\left(k_{p}^{2} + k_{z}^{2}\right)
\text{sinch}\left(k_{z}t\right)+1\right)}}.
\end{align}
The above basis waves are also valid for periodic films. 
The only change required is to replace the continuous index 
$\textbf{k}_{p}$ by its appropriate discrete counterpart. 

\subsection{Inverse Design Procedure}

All computational geometries were discovered following a standard
topology optimization algorithm~\cite{molesky2018inverse}, where each
pixel (susceptibility value) within the bounding sphere is considered
as an independent design parameter, based on the method of moving
asymptotes~\cite{svanberg2002class}. To begin, each pixel (index by
$\textbf{x}$) is allowed to linearly explore a continuous spectrum of
susceptibility values varying between the background
$\chi_{\textbf{x}} = 0$ and the stated material $\chi_{\textbf{x}} =
\chi$. The resulting parameter values are then iteratively used as
starting points for new optimizations subject to increasingly strict
regularization filters, as described by Jensen and
Sigmund~\cite{jensen2011topology}, in order to enforce binarization
(each pixel must correspond to either the material $\chi$ or
background).

The primary challenge of applying this approach to maximize the 
scattered power from a planewave incident on a three-dimensional compact
object, \eqref{fSct}, lies in the computational cost of  
$P^{\mathrm{sct}}_{\mathrm{flx}}$, which is typically evaluated several 
thousand times during the course of any one optimization. Without 
simplification, each evaluation requires a new solution of 
Maxwell's equations in three-dimensional space. 
To avoid this otherwise significant computational challenge, 
the results presented in Sec.~\ref{applications} exploit a 
fluctuating--volume current (FVC) formulation of electromagnetic 
scattering~\cite{jin2016temperature,molesky2020fundamental} 
and the efficiency of stochastic singular value decomposition. 
(Descriptions of the various subroutines entering into this 
program are given in~\cite{halko2011finding,martinsson2011randomized,
polimeridis2015fluctuating}.)
This approach both limits the spatial solution domain to 
the bounding sphere, and allows the central matrix-vector multiplication, 
$\mathbb{G}^{\text{0}} \left|\textbf{J}^{g}\right>$,
to be accelerated via the fast-Fourier transform.

More precisely, optimization is carried out using the first form of the 
scattered power quoted in \eqref{fSct},
\begin{equation}
P^{\mathrm{sct}}_{\mathrm{flx}} = \frac{k}{2Z}
\left<\textbf{T}\right|\asym{\mathbb{G}^{\text{0}}}\left|\textbf{T}\right>
\label{eqVIE}
\end{equation}
with $\left|\textbf{T}\right> =
\mathbb{T}_{ss}\left|\textbf{E}^{i}\right>$, as an objective.  To
compute the real number associated with this form for any particular
structure, one inverse solve (sped up by applying a fast-Fourier
transform) is needed to determine $\left|\textbf{T}\right>$, which
from \eqref{initFlux} is proportional to the generated current
$\left|\textbf{J}^{g}\right>$.  The gradient of
$P^{\mathrm{sct}}_{\mathrm{flx}}$ with respect to the susceptibility
degrees of freedom then follows from the defining relation $\mathbb{T}
= \left(\mathbb{V}^{-1}-\mathbb{G}^{\text{0}}\right)$.  (At this step,
a nonzero but arbitrarily small scattering potential is assumed to
exist at all points within the domain so that $\mathbb{V}^{-1}$ is
well defined throughout the ball.)  Parametrizing the susceptibility
at each pixel by the single material linear equation
$\chi_{\textbf{x}} = t_{\textbf{x}}~\chi$, with
$t_{\textbf{x}}\in\left(0,1\right]$,
$$\partial_{t_{\textbf{x}}}\mathbb{T} = 
\chi~\mathbb{T}\mathbb{V}^{-1}
\delta_{\textbf{x},\textbf{x}}\mathbb{V}^{-1}\mathbb{T} = 
\chi~ \mathbb{W}\delta_{\textbf{x},\textbf{x}}\mathbb{W}.$$ 
Using this result, the total gradient 
(with well defined $t_{\textbf{x}}\rightarrow 0$ limits) is then
\begin{align}
\frac{\partial P^{\mathrm{sct}}_{\mathrm{flx}}}{\partial t_{\textbf{x}}} 
= -\frac{k}{Z} \left<\textbf{T}\right|\sym{\delta_{\textbf{x},\textbf{x}}
\mathbb{W}\asym{\mathbb{G}^{\text{0}}}}
\left|\textbf{T}\right>
\label{derVIE}
\end{align}
Hence, only a single additional solve is required to obtain complete 
gradient information for the optimization objective. 
For the periodic films, the optimization is carried out using the 
rigorous coupled wave approach recently presented in 
Ref.~\cite{jin2020inverse}.
\bibliography{esaLib}
\end{document}